\documentclass[twocolumn]{aastex631}

\newcommand{\au}{{\rm{AU}}}

\newcommand{\Rp}{{r}}
\newcommand{\Mp}{{m}}
\newcommand{\Ms}{{M_{\rm{*}}}}

\newcommand{\Mpone}{{m_{\rm{1}}}}
\newcommand{\Mptwo}{{m_{\rm{2}}}}
\newcommand{\Mptot}{{m_{\rm{tot}}}}
\newcommand{\Rpone}{{r_{\rm{1}}}}
\newcommand{\Rptwo}{{r_{\rm{2}}}}
\newcommand{\Rcomb}{{r_{\Mpone + \Mptwo}}}
\newcommand{\aone}{{a_{\rm{1}}}}
\newcommand{\atwo}{{a_{\rm{2}}}}

\newcommand{\dMp}{{\Delta \Mp}}

\newcommand{\dMpbydelt}{{|\Delta \Mp / \Delta t|}}

\newcommand{\dMponebyMpone}{{\Delta \Mpone / m_{\rm{1i}}}}
\newcommand{\dMptwobyMptwo}{{\Delta \Mptwo / m_{\rm{2i}}}}
\newcommand{\dMptotbyMptot}{{\Delta \Mptot / m_{\rm{tot,i}}}}

\newcommand{\dMgonebyMgone}{{\Delta m_{\rm{gas,1}} / m_{\rm{gas,1i}}}}
\newcommand{\dMgtwobyMgtwo}{{\Delta m_{\rm{gas,2}} / m_{\rm{gas,2i}}}}
\newcommand{\dMgtotbyMgtot}{{\Delta m_{\rm{gas,tot}} / m_{\rm{gas,tot,i}}}}

\newcommand{\done}{{d_{\rm{1}}}}
\newcommand{\dtwo}{{d_{\rm{2}}}}
\newcommand{\Lone}{{L_{\rm{1}}}}

\newcommand{\ecc}{{e}}
\newcommand{\inc}{{i}}

\newcommand{\Rc}{{r_{\rm{c}}}}
\newcommand{\Rm}{{r_{\rm{m}}}}

\newcommand{\Ts}{{T_{\rm{s}}}}
\newcommand{\Ps}{{P_{\rm{s}}}}

\newcommand{\bimpact}{{b^{\prime}}}
\newcommand{\vimpact}{{v_{\rm{im}}}}
\newcommand{\vesc}{{v_{\rm{esc}}}}
\newcommand{\vmesc}{{v_{\rm{m,esc}}}}
\newcommand{\Eimpact}{{E_{\rm{im}}}}

\newcommand{\bcrit}{{b_{\rm{c}}^{\prime}}}
\newcommand{\bcritvesc}{{b_{\rm{c,1}}^{\prime}}}
\newcommand{\bcritvesctwo}{{b_{\rm{c,2.5}}^{\prime}}}

\newcommand{\gcc}{{{\rm g\ cm^{-3}}}}

\begin{document}

\title{Outcomes of Sub-Neptune Collisions}
\shorttitle{Outcomes of Sub-Neptune Collisions}
\shortauthors{Ghosh, Chatterjee \& Lombardi 2023}

\author[0000-0002-3103-2000]{Tuhin Ghosh}
\affiliation{Department of Astronomy and Astrophysics, Tata Institute of Fundamental Research, Homi Bhabha Road, Navy Nagar, Colaba, Mumbai, 400005, India}
\email{tghosh.astro@gmail.com}

\author[0000-0002-3680-2684]{Sourav Chatterjee}
\affiliation{Department of Astronomy and Astrophysics, Tata Institute of Fundamental Research, Homi Bhabha Road, Navy Nagar, Colaba, Mumbai, 400005, India}
\email{souravchatterjee.tifr@gmail.com}

\author[0000-0002-7444-7599]{James C. Lombardi, Jr.}
\affiliation{Department of Physics, Allegheny College, Meadville, Pennsylvania 16335, USA}
\email{jamie.lombardi@allegheny.edu}
 

\begin{abstract}
Observed high multiplicity planetary systems are often tightly packed. Numerical studies indicate that such systems are susceptible to dynamical instabilities. Dynamical instabilities in close-in tightly packed systems, similar to those found in abundance by Kepler, often lead to planet-planet collisions. For sub-Neptunes, the dominant type of observed exoplanets, the planetary mass is concentrated in a rocky core, but the volume is dominated by a low-density gaseous envelope. For these, using the traditional perfect merger assumption (also known as the `sticky-sphere' approximation) to resolve collisions is questionable. Using both N-body integration and smoothed-particle hydrodynamics, we have simulated sub-Neptune collisions for a wide range in realistic kinematic properties such as impact parameters ($\bimpact$) and impact velocities ($\vimpact$) to study the possible outcomes in detail. We find that the majority of the collisions with kinematic properties similar to what is expected from dynamical instabilities in multiplanet systems may not lead to mergers of sub-Neptunes. Instead, both sub-Neptunes survive the encounter, often with significant atmosphere loss. When mergers do occur, they can involve significant mass loss and can sometimes lead to complete disruption of one or both planets. Sub-Neptunes merge or disrupt if $\bimpact<\bcrit$, a critical value dependent on $\vimpact/\vesc$, where $\vesc$ is the escape velocity from the surface of the hypothetical merged planet assuming perfect merger. For $\vimpact/\vesc\lesssim2.5$, $\bcrit\propto(\vimpact/\vesc)^{-2}$, and collisions with $\bimpact<\bcrit$ typically leads to mergers. On the other hand, for $\vimpact/\vesc\gtrsim2.5$, $\bcrit\propto\vimpact/\vesc$, and the collisions with $\bimpact<\bcrit$ can result in complete destruction of one or both sub-Neptunes.

\end{abstract}



\section{Introduction}\label{sec:intro}

In the last few decades, we have detected over $5,000$ exoplanets in more than $3,500$ planetary systems \citep[NASA Exoplanet Archive\footnote{http://exoplanetarchive.ipac.caltech.edu},][]{2013Akeson_NEA}, revealing a diverse population of planets with little to no resemblance to the solar system. Among these, most abundant are the short-period ($<100$ days) super-Earths and sub-Neptunes, that have no solar system equivalent. Apart from the sheer number of planet discoveries, we now have a wealth of data on hundreds of multi-planet systems \citep[e.g.,][]{2011_Lissauer,2014_Fabrycky}. These systems are often dynamically packed and close to their stability limit \citep[e.g.,][]{Deck_2012, 2013Fang}. Numerous studies regarding the dynamical evolution and stability of such multi-planet systems have suggested that dynamical instabilities play a significant role in the final planet assembly and likely shaped the orbital architectures of planetary systems we observe today \citep[e.g.,][]{1996_Rasio_Ford, Chatteerjee_2008, 2008Juric_Tremaine, 2015PuWu, Volk_2015, 2017Izidoro, Frelikh_2019, 2020_Anderson, Goldberg_2022, 2023_Lammers, 2023_Ghosh_PPScattering}.

Dynamical instability excites the planetary orbits which lead to close encounters between planets, eventually resulting in planet-planet collisions or ejections. In close-in planetary systems, planet-planet collisions are more common \citep[e.g.,][]{2023_Ghosh_PPScattering}. In the $N$-body dynamical models of planetary systems, the collisions are generally resolved using the perfect merger approximation (commonly referred to as the `sticky-sphere' approximation), where the colliding planets are merged into a single body, conserving mass and linear momentum. This simple approximation does not consider the detailed kinematics of different collisions and treats all collisions as perfect inelastic mergers, irrespective of the impact parameter and the relative velocity between the colliding bodies. In principle, depending on the kinematics of the specific collision, the planets may not always merge, such as in high-velocity collisions with impact parameters comparable to their radii \citep[e.g.,][]{2004_Agnor_Asphaug, 2012_Leinhardt}.
This is particularly relevant for sub-Neptune-type planets. The structure of these planets is such that most of the mass is concentrated in the core, while a low-density gaseous envelope constitutes most of its volume \citep[e.g.,][]{Hadden_2014, 2014_Lopez_Fortney, Weiss_2014, 2015_Rogers, 2015_Wolfgang_Lopez}. Hence, the perfect merger approximation, ubiquitous in $N$-body dynamical integrations, is questionable and may not be appropriate in many cases. Tackling these issues requires thorough numerical experiments involving proper three-dimensional multi-layered planet models and detailed hydrodynamic simulations of collisions between planets in a variety of collision scenarios.

Several studies of planetary collisions using numerical hydrodynamic simulations were conducted in the past, primarily to understand moon formation \citep[e.g.,][]{1986_Benz,2001_Canup,CANUP_2013, Kegerreis_2022}. Even from the early days, it was clear that not all collisions between protoplanets result in perfect mergers \citep[e.g.,][]{2004_Agnor_Asphaug, 2006_Asphaug_Nature, 2012_Genda, 2012_Leinhardt}. Most of these early studies were motivated by the giant impact phase of our solar system and considered collisions between sub-Earth-sized protoplanets or planetesimals. Based on these simulations, several scaling laws and even machine learning models were developed to predict the outcome of such collisions across a wide range of initial conditions \citep[e.g.,][]{2012_Leinhardt, 2019_Cambioni, 2020_Emsenhuber}. Motivated by the recent discoveries of super-Earths and sub-Neptunes, \citet{2009_Marcus, 2010_Marcus} studied collisions between rocky super-Earths. Later studies included the presence of an H/He atmosphere; however in most cases, they considered collisions with an atmosphere-less projectile \citep[e.g.,][]{Liu_2015, 2018_Kegerreis, 2020_Kegerreis, 2020_Denman, 2022_Denman} with the notable exception of \citet{2017_Hwang, 2018_Hwang}, who studied highly grazing collisions of multi-layered sub-Neptune type planets. In recent years collisions between massive Jupiter-like planets were also investigated \citep{2020_Li}.

In this study, we revisit the problem of sub-Neptune collisions. We use smoothed particle hydrodynamics \citep[SPH,][]{1977_Gingold_Monaghan,1977_Lucy} to make 3D models of differentiated, three-component planets and conduct detailed numerical experiments of planet-planet collisions, varying the initial conditions, to gain insight into the outcomes of such violent events. SPH is a widely used meshless hydrodynamic scheme, where the material is modeled by a finite number of fluid elements or `SPH particles' \citep[see][for reviews]{Monaghan_2005, 2010_Springel, PRICE2012}. While \citet{2018_Hwang} were able to simulate grazing collisions of sub-Neptunes, where only the gaseous envelopes interact, our setup can simulate collisions of all possible impact parameters, including head-on collisions involving the planetary cores. This improvement is made possible by using appropriate equations of state (EOSs) for the planetary materials and a better treatment of SPH particles at the boundaries between core, mantle, and atmosphere (see \autoref{subsec:setup/hydro}).

The rest of the paper is structured as follows. In \autoref{sec:setup}, we discuss the chosen initial conditions and describe the dynamical integrations to generate planet-planet collisions. We also present our numerical setup for creating 3D SPH planet models and performing the subsequent detailed hydrodynamic simulations of planet-planet collisions. In \autoref{sec:results}, we present our key results. We summarise and conclude in \autoref{sec:summary}.

\section{Numerical Setup}\label{sec:setup}

In this study, we consider two sub-Neptune planets with masses $\Mpone = 4.51 M_{\earth}$, $\Mptwo = 7.55 M_{\earth}$ and radii $\Rpone = 3.56 R_{\earth}$, $\Rptwo = 3.69 R_{\earth}$ (listed in \autoref{tab:planet_profiles}), orbiting around a $1.03 M_{\sun}$ star. In our setup, we first initialize the planets in unstable orbits that would eventually lead to collisions. We use the pre-collision orbital properties to conduct detailed hydrodynamical simulations using SPH to determine the outcome of the collisions.

\subsection{Pre-Collision Orbits}\label{subsec:setup/pre_coll}
To generate planet-planet collisions, we create $N$-body models of the chosen planets in orbits around their host star. The inner planet's semi-major axis is set to $\aone = 0.115 \au$. The outer planet is placed at a period ratio of $1.3$, corresponding to $\atwo = 0.137 \au$. The planetary eccentricities {$\ecc$} are randomly chosen from a uniform distribution within $0 < \ecc < 0.9$. The inclinations ($\inc$) are randomly drawn from the Rayleigh distributions with $ \bar{\inc} = 0.024$ \citep{2016_Xie}. We integrate these systems with the \texttt{IAS15} integrator \citep{2015_IAS15_REB} implemented in the \texttt{REBOUND} \citep{2012_REB} simulation package. A collision is considered when the distance between the planets is less than $\Rpone + \Rptwo$. We stop the integrations whenever a collision is detected. Otherwise, we integrate till a predetermined stopping time ($\sim 16000$ years, $\gtrsim4\times10^5$ times the inner planet's initial orbital period). In the event of a collision, we store a snapshot of the planetary orbits before the collision when the distance between the centres of mass for the planets is $5 \times (\Rpone + \Rptwo)$. These snapshots would later serve as the starting points of the SPH calculations. We simulate over 30000 of these systems to densely populate the parameter space in the impact parameters ($b$, the perpendicular distance between the velocity vectors of the two planets) and impact velocities ($\vimpact$) of planet-planet collisions. We measure $b$ and $\vimpact$ when their surfaces make contact, i.e., the distance between the centres of mass of the planets is $\Rpone + \Rptwo$. Throughout this work, we represent the impact parameter and impact velocity as dimensionless ratios $\bimpact \equiv b/(\Rpone + \Rptwo)$ and $\vimpact/\vesc$ where, $\vesc = \sqrt{2G(\Mpone + \Mptwo)/\Rcomb}$, $G$ is the gravitational constant, $\Rcomb = (\Rpone^3 + \Rptwo^3)^{1/3}$ is the radius of a planet with mass $\Mpone + \Mptwo$ and the same average density of the colliding planets.
\footnote{Note that another commonly used unit for measuring $\vimpact$ is the so called mutual escape velocity, defined as $\vmesc = \sqrt{2G(\Mpone + \Mptwo)/ (\Rpone + \Rptwo)}$. For the planets considered in our study, this differs from our chosen unit by a constant factor, $\vesc = 1.26 \times \vmesc$.}
The dimensionless impact parameter ($\bimpact$) is related to the impact angle ($\theta$, the angle between the line connecting the centers of the two bodies and their relative velocity vector) as $\bimpact = \sin{\theta}$.

\begin{figure}[htb!]
\includegraphics[width=\columnwidth]{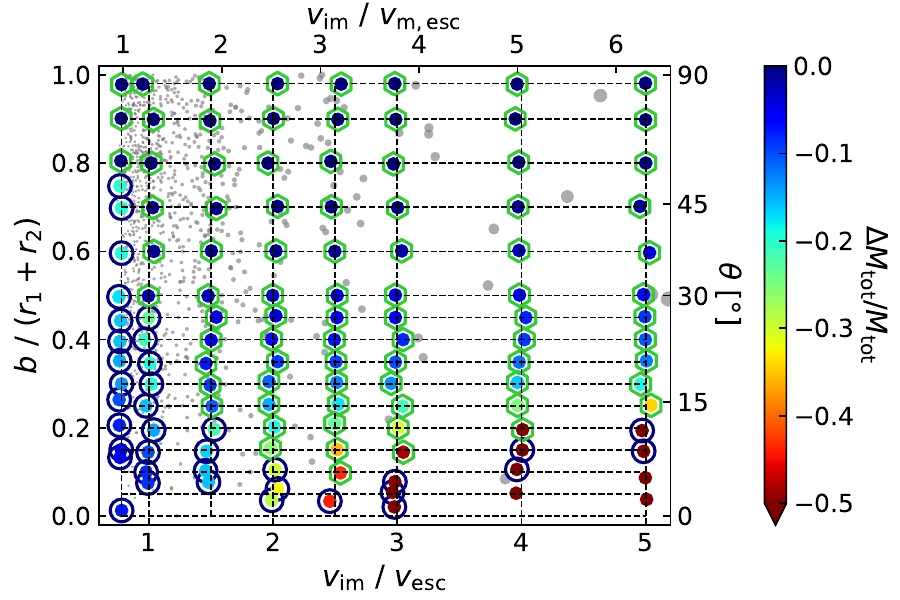}
\caption{The 2D grid of simulations considered in our study in impact parameter ($\bimpact$) and impact velocity ($\vimpact$) at the moment of contact. The colored dots represent the simulations conducted in this study. The colors of the individual crosses reflect the overall mass loss during the encounter. The blue circles (green hexagons) enclosing the crosses highlight the simulations where one (or both) planet(s) survive. Collisions resulting in the destruction of both planets are shown without any enclosure. The grey dots represent the collisions produced in typical planet-planet scattering experiments (adapted from ensemble \texttt{n8-e040-i024} of \citet{2023_Ghosh_PPScattering}). The dot sizes are proportional to the impact energy.}
\label{fig:grid}
\end{figure}

We construct a two-dimensional grid using the $N$-body experiments described earlier, summarized in \autoref{fig:grid} and \autoref{tab:all_sims}. The extent and spacing between the grid points are motivated by the planet-planet collisions found in $N$-body scattering experiments conducted in \citet[][adapted from their ensemble \texttt{n8-e040-i024}, indicated by the background grey dots in \autoref{fig:grid} for reference]{2023_Ghosh_PPScattering}.\footnote{Note that the planet-planet scattering experiments described in \citet{2023_Ghosh_PPScattering} involve sub-Neptunes of various sizes, whereas our study focuses on two specific sub-Neptunes. Despite this difference, we believe that the collision kinematics found in their scattering experiments provide valuable insights into the parameter space applicable to sub-Neptune collisions.} We first construct the desired grid, and then at each grid point, we assign the nearest collision (in $\bimpact-\vimpact/\vesc$ parameter space) from the collisions generated in our $N$-body experiments. Since our $N$-body collision sample is finite, populating some grid points, particularly those at  $\bimpact \sim 0$, is challenging and sometimes results in vacant grid points (see \autoref{fig:grid}).
%
%

\subsection{Hydrodynamic Simulations}\label{subsec:setup/hydro}

Hydrodynamic simulations of planet-planet collisions were carried out using the SPH code \texttt{StarSmasher} \citep{1991_Rasio_Thesis,2010_Gaburov,2018_Gaburov_Code}. This code implements an artificial viscosity prescription coupled with a Balsara Switch \citep{1995_BALSARA} to avoid unphysical inter-particle penetration \citep[described in][]{2015_Hwang_Balsara}. For our calculations, we use the Wendland C4 smoothing kernel \citep{Wendland1995PiecewisePP}. In typical hydrodynamical simulations of planet-planet collision, the central star is often neglected \citep[e.g.,][]{2012_Leinhardt, 2020_Kegerreis, 2022_Denman}. However, the central star can significantly impact the collision outcome, especially in collisions involving multiple encounters \citep{2019_Emsenhuber_Asphaug}. In this study, we include the central star in our calculations and treat the star as a non-interacting point-mass particle, exerting only gravitational influence over the planets. If a particle approaches within one solar radius of the star, it is considered accreted into the star, conserving mass and linear momentum.

We model the planets as differentiated bodies consisting of three layers - an iron core, a rocky mantle, and a hydrogen-helium (H-He) envelope. We implemented the widely used \citet{1962_Tillotson} iron and granite EOS (parameters are taken from Table A1 of \citet{2017_Reinhardt}) to model the core and mantle materials respectively. For the H-He envelope, we use the \citet{1980_HM80_EOS} EOS. In order to obtain the equilibrium planet profile, we utilize the code \texttt{WoMa} \citep{2020_WoMa}. We assume that $15\%$ of the total planet mass is in the H-He envelope, and the rest is in the iron core and rocky mantle in $30:70$ ratio. We truncate the planet profile at a negligible density of $0.01 \gcc$ and consider this as the planet's surface. We fix the surface temperature ($\Ts$) to our desired value depending on the insolation flux and assume an adiabatic temperature profile for the atmosphere. The core and the mantle are assumed to be isothermal. For simplicity, we assume that the planets have no initial spin. The parameters used to generate the equilibrium planet profiles including the outer radii of core ($\Rc$), mantle ($\Rm$), and the planet, and the pressure at the outermost surface layer ($\Ps$) of the planets are provided in \autoref{tab:planet_profiles}.

After obtaining the planet profiles, we place the SPH particles accordingly in concentric spherical shells using the stretched equal-area method \citep{2019_Kegerreis_SEAGEN}. The majority of the previous studies \citep[e.g.,][]{2020_Denman,2022_Denman,2020_Kegerreis,2020_Kegerreis_APJL} have employed equal mass SPH particles with the most prominent exception of \citet{2017_Hwang, 2018_Hwang}. However, opting for equal-mass SPH particles would lead to a concentration of most particles in the dense inner region of the planet, while the outer low-density H-He envelope, which constitutes most of the planet's volume, would have a significantly lower resolution. We tackle this problem by utilizing \texttt{StarSmasher}'s ability to handle unequal mass particles \citep{2010_Gaburov} and setting the particle mass proportional to the density of its position inside the planet. This choice ensures a relatively uniform number density of SPH particles throughout the planet's radial profile. Nevertheless, to prevent the accumulation of very low-mass particles in the outermost, least-dense layers, we limit the minimum mass for SPH particles to $1/1000$th of the most massive SPH particle.
Although particle mass differences as large as 1000:1 or more are often utilized in studies of stellar tidal disruptions \citep[e.g.,][]{2011_Fabio, 2023_Kiroglu}, these studies typically treat bodies as ideal gasses and interactions are more gradual. In contrast, our focus is on hypervelocity collisions between differentiated planetary bodies with density discontinuities between different layers and modeled using condensed-matter equations of state, necessitating caution in the application of these methods.\footnote{A similar approach was taken by \citet{2017_Hwang,2018_Hwang}.}

Our models have a resolution of $\sim10^5$ particles per planet. Note that some studies using equal-mass SPH particles advocate for even higher resolution in modeling planetary collisions \citep[e.g.,][]{2019_Kegerreis_SEAGEN}. However, we find that results from representative higher-resolution simulations remain well within the expected dispersion (caused by tiny variations in initial conditions) in those modeled with our fiducial resolution. See \autoref{app:convergence} for a more detailed discussion on the numerical convergence.

This SPH setup cannot yet be simulated because of the steep density discontinuities at the core-mantle and mantle-atmosphere boundaries. The sharp density discontinuities in the material boundaries can not be represented in the standard SPH framework because of the inherent smoothing of the density. Thus, if unaddressed, density-smoothing breaks the pressure continuity near these boundaries, resulting in high pressure gradients that lead to large artificial forces on the nearby SPH particles \citep{2007_Woolfson, 2022_Ruiz-Bonilla}. Previous attempts to address this complication include changing the formulation of SPH \citep[e.g.,][]{2008_PRICE, Saitoh_2013, 2016_Hosono}, implementing corrections for the SPH density \citep[e.g.,][]{2007_Woolfson, 2020_Reinhardt, 2022_Ruiz-Bonilla} and using the so-called "mixed composition" particles near the boundaries \citep[e.g.,][]{2017_Hwang, 2018_Hwang}. In our study, we adopt the prescription of mixed-composition particles outlined by \citet{2017_Hwang}. We replace the particles affected by density-smoothing with mixed composition particles. These particles are assumed to contain two different materials of different densities at pressure equilibrium with each other, such that the combined particle has the same density as the smoothed SPH density while conserving mass and preserving pressure continuity. For more details on the algorithms used to set up the mixed-composition particles, see Section A3 of \citet{2017_Hwang}, noting that we have updated the EOSs. These mixed composition particles are needed only to create stable SPH models of differentiated planets initially and have no role when they are outside the planet. Therefore, we allow these particles to revert to their initial pure compositions whenever these particles are outside a planet-like environment (when surrounding density is $<10^{-3}\gcc$) at any point during the simulations to avoid complications in the pressure calculation algorithm specifically in very low-density regions.

\begin{figure}[htb!]
\includegraphics[width=\columnwidth]{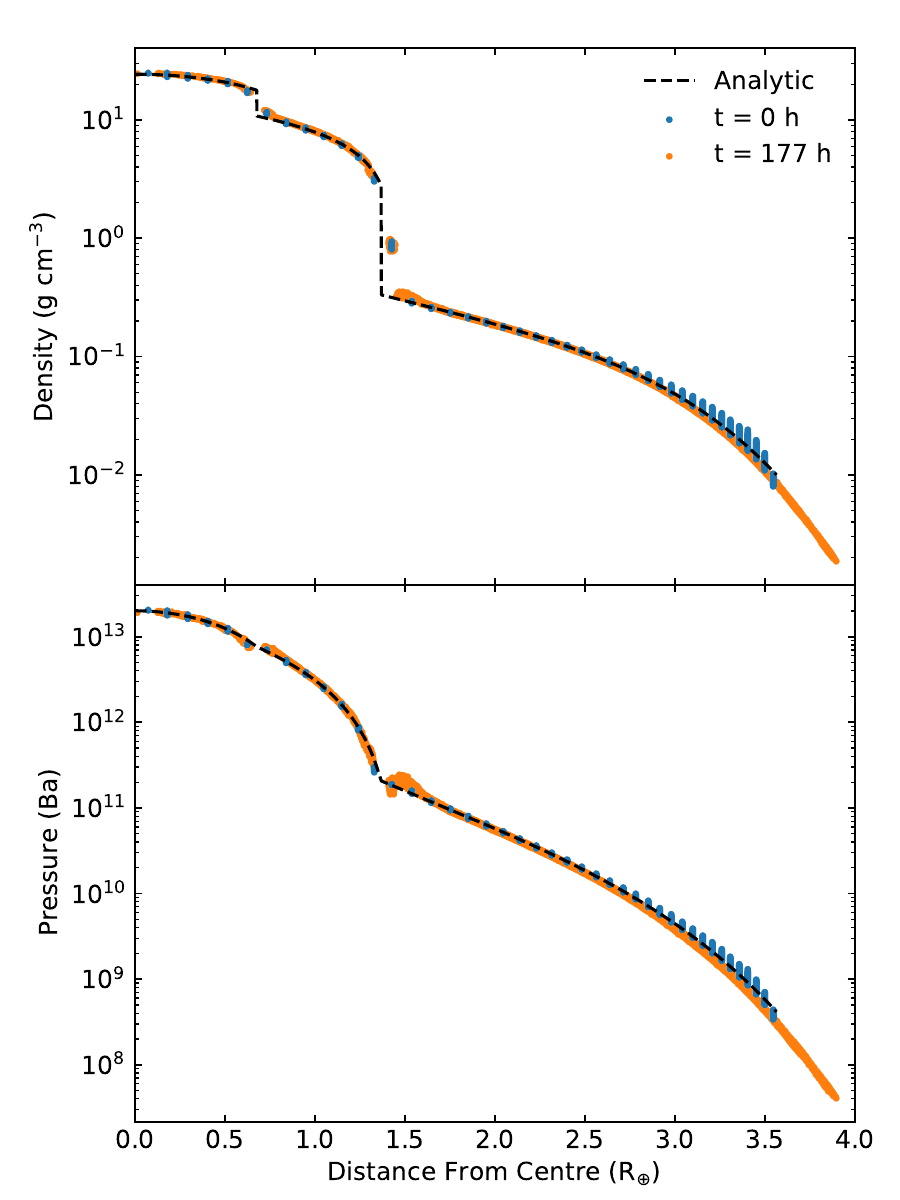}
\caption{Density (upper panel) and pressure (lower panel) profiles of Planet-1 (\autoref{tab:planet_profiles}). The black dashed line shows the analytic profile of the planet in hydrostatic equilibrium. The initial profiles of the SPH planet are shown in blue, and the same after relaxing the SPH planet in isolation for $177$ hours, is shown in orange.
}
\label{fig:planet1_profile}
\end{figure}

\autoref{fig:planet1_profile} shows the radial density and pressure profiles of Planet-1 as an example. Our SPH planet model is consistent with the analytic profile from \texttt{WoMa} even after the planet is relaxed in isolation for 177 hours, indicating that our SPH planet models are adequately stable. Note that there is some leakage of particles beyond the supplied outer edge of the planet during relaxation. This is likely related to the choice of the minimum density in the supplied planet profile. Had we chosen a lower density cutoff, the SPH density would have been even more consistent. Because the mass of this extended low-density layer is negligible, it is not expected to affect our simulations. The final snapshots from these relaxation runs are taken and initialized in pre-collision orbits obtained in \autoref{subsec:setup/pre_coll}, around the point-mass host star. We now start our hydrodynamic simulation with both planets in a collision course and run it for a minimum duration of 44.3 hours. The eventual integration stopping time depends on the details of the collision and is discussed in \autoref{subsec:setup/sim_stopping}.

\begin{deluxetable}{ccccccc}
\tablecaption{Initial Planet Properties
\label{tab:planet_profiles}}
\tablewidth{0pt}
\tablehead{
    \colhead{} & 
    \colhead{$\Mp/M_{\earth}$} & 
    \colhead{$\Rp/R_{\earth}$} & 
    \colhead{$\Rc/R_{\earth}$} &
    \colhead{$\Rm/R_{\earth}$} & 
    \colhead{$\Ts/K$} &
    \colhead{$\Ps/Ba$}
    }

\startdata
    Planet-1 & $4.51$ & $3.56$ & $0.68$  & $1.37$  & $1091$ &  $4.15\times10^{8}$ \\
    Planet-2 & $7.55$ & $3.69$ & $0.75$  & $1.51$  & $1030$ &  $3.93\times10^{8}$ \\
\enddata

\tablecomments{Initial properties such as mass ($\Mp$), total radius ($\Rp$), outer radius of the core ($\Rc$), outer radius of the mantle($\Rm$), surface temperature ($\Ts$) and pressure ($\Ps$) of the two planets considered in this study.} 
\end{deluxetable}
%

\subsection{Identifying Post-Collision Planets}\label{subsec:setup/post_coll}

After conducting the hydrodynamic calculations, we are left with a distribution of fluid particles. However, this distribution does not immediately tell us the number of planets remaining in the simulation post-collision and which fluid particles constitute a planet. To identify the planetary bodies, we first use the \texttt{DBSCAN} \citep{DBSCAN} algorithm to identify clusters of core and mantle particles in the final snapshot of the hydrodynamic calculation. We treat these clusters as the initial planet `seed' candidates based on which we would ultimately reconstruct the planets emrging out of the collision. Subsequently, we calculate the Jacobi constant for all the fluid particles corresponding to each of these planet candidates as \citep{murray_dermott_2000}
\begin{equation}
\label{eqn:jacobi_constant}
C_{\rm{J}} = \frac{x^2 + y^2}{r^2} + \frac{2r}{\Ms+\Mp}\left( \frac{\Ms}{\done} + \frac{\Mp}{\dtwo} \right) - \frac{r v^2}{G(\Ms+\Mp)}
\end{equation}
where $\Ms$ and $\Mp$ are the masses of the host star and the planet candidate, $r$ is the distance between the star and planet candidate, $\done$ and $\dtwo$ are the separations of the particle from the star and the planet candidate, ($x$, $y$) and $v$ represent the position co-ordinates and the velocity of the fluid particle in the co-rotating frame of the planet candidate and the star with origin at the center of mass. We then calculate the Jacobi constant at Lagrange point, $\Lone$ of the planet candidate, assuming a circular orbit \citep{murray_dermott_2000},
\begin{equation}
\label{eqn:jacobi_constant_at_L1}
C_{\Lone} \approx 3 + 3^{4/3} \mu_{2}^{2/3} -10 \mu_{2}/3
\end{equation}
where $\mu_{2} = \Mp/(\Ms+\Mp)$.

We assign the fluid particles to the planet candidate if they are located within the candidate's Hill sphere and satisfy the criterion $C_{\rm{J}}  > C_{\Lone}$. Next, we update the planet candidates' mass, position, and velocity, taking into account the newly assigned particles. We then recalculate the Jacobi constants for the updated planet candidates and assign additional particles bound to these updated candidates. To ensure a reliable reconstruction of planetary bodies from these fluid particles, we repeat this process multiple times iteratively using the updated planet candidate from the previous iteration as the seed for the next iteration until convergence \footnote{Additionally, we exclude any planet candidates less massive than $0.1 M_{\earth}$ to prevent misidentifying very small clusters of fluid particles as planets.}.

\subsection{Determining Runtime of Simulations}\label{subsec:setup/sim_stopping}

After running the simulations for at least $44.3$ hours, we inspect the last snapshot and determine the number of planets still remaining in the simulation (see \autoref{subsec:setup/post_coll} for details). If two planets remain with overlapping Hill spheres, we continue the simulation until their separation exceeds the combined size of their Hill spheres. This was needed for only one low $\vimpact/\vesc = 0.78$ collision. Note that if we integrate such collisions further, there is a possibility that in their new orbits they might have another collision. However, in our framework, this would be considered a separate collision between planets with different initial structures and kinematics given by those of the collision remnants we have simulated. On the other hand, if only one planet survives, it may take a long time to settle, and our planet identification algorithm (\autoref{subsec:setup/post_coll}) may give us different masses if applied to snapshots taken at different times. As time progresses, the planet gradually stabilizes, and the results from the planet identification algorithm converge. We track this convergence using the difference in planet mass ($\dMp$) as calculated by our algorithm over time, and extend the simulation if necessary until $\dMpbydelt$ falls below the threshold of $0.1\,M_{\earth}\rm{day}^{-1}$ or until we reach a predefined maximum duration of 332 hours (equivalent to the initial orbital period of the inner planet), imposed due to computational limitations. The chosen simulation stopping times for all the simulations are listed in \autoref{tab:all_sims}. Note that for the majority of our simulations, $\dMpbydelt$ is significantly below our specified threshold; the 16th percentile, median, and 84th percentiles for $\dMpbydelt$ are $2.1\times 10^{-4}$, $4.1\times 10^{-3}$, and $3.5\times 10^{-2}$ $M_{\earth}\rm{day}^{-1}$, respectively.

\section{Results}\label{sec:results}

\begin{figure*}[htb!]
\includegraphics[width=\textwidth]{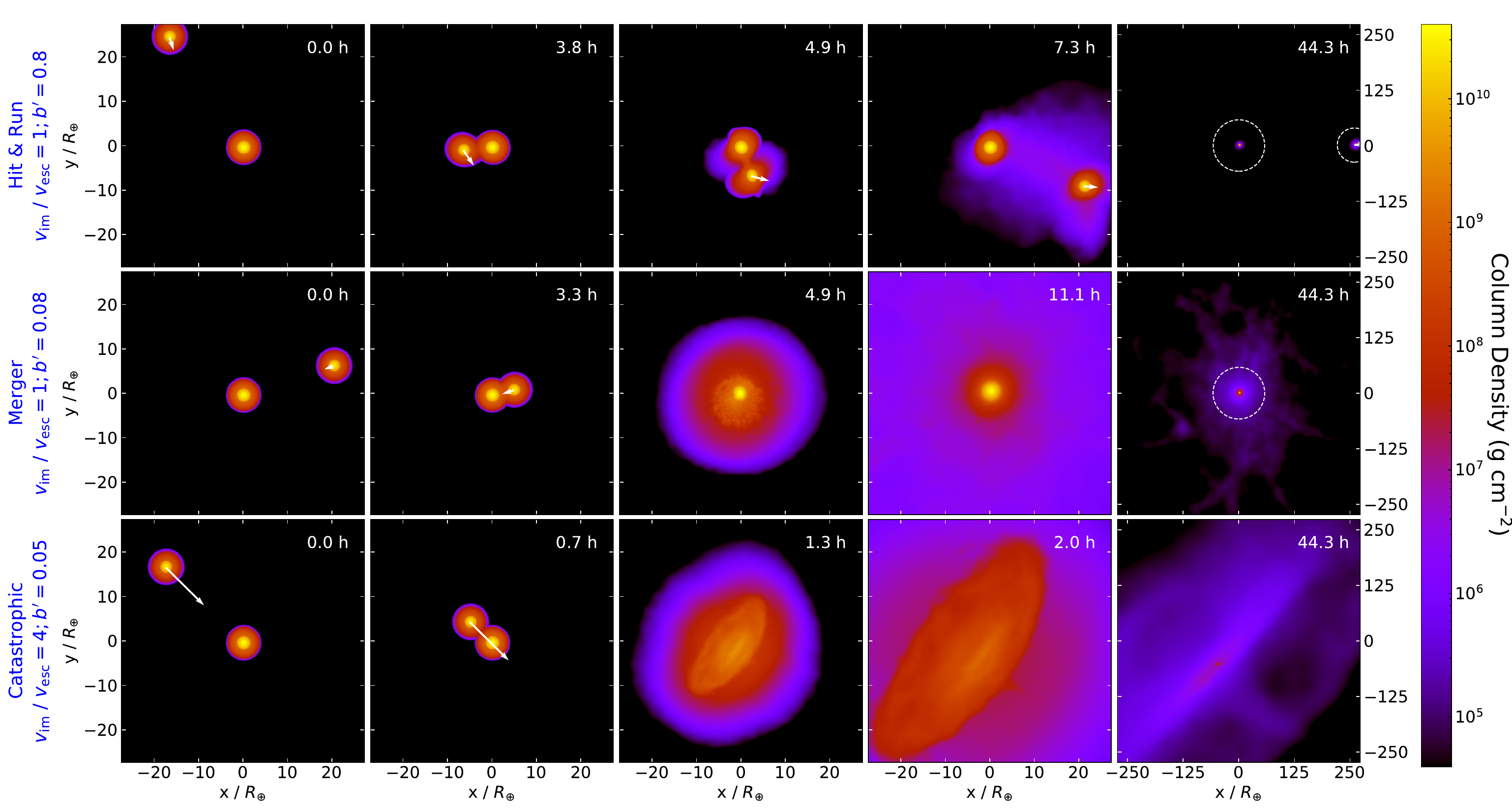}
\caption{Snapshots showing the different stages of the encounter for the three major types of outcomes. In each panel, the reference frame is centered at the position of the most massive SPH particle, initially situated at the center of the more massive planet, and the colors represent the column density. The white arrows on the less massive planet represent its velocity relative to the more massive planet, with arrow length proportional to the velocity. The Hill spheres of the planets are shown by the white dashed circles on the rightmost panels. The top (middle) row shows an example of \texttt{Hit \& Run} (\texttt{Merger}) encounter with $\vimpact \sim \vesc$ and $\bimpact=0.8$ ($\bimpact=0.08$). The bottom row shows the snapshots of a highly energetic ($\vimpact/\vesc \approx 4$) encounter with low impact parameter ($\bimpact = 0.05$) that results in the destruction of both the planets, categorized as a \texttt{Catastrophic} collision.}
\label{fig:snapshots}
\end{figure*}

The outcomes of our collision simulations are summarized in \autoref{fig:grid} and \autoref{tab:all_sims}. We find that the outcomes of these collisions can be categorized into three major groups depending on the remaining number of planets, namely a) \texttt{Hit \& Run} where both planets survive the encounter, b) \texttt{Merger} where the planets merge to form a single planet, and c) \texttt{Catastrophic}, violent collisions resulting in the destruction of one or both planets. \autoref{fig:snapshots} shows examples of the time evolution of our hydrodynamic calculations, illustrating the different stages for each type of outcome.

\begin{figure}[htb!]
\includegraphics[width=\columnwidth]{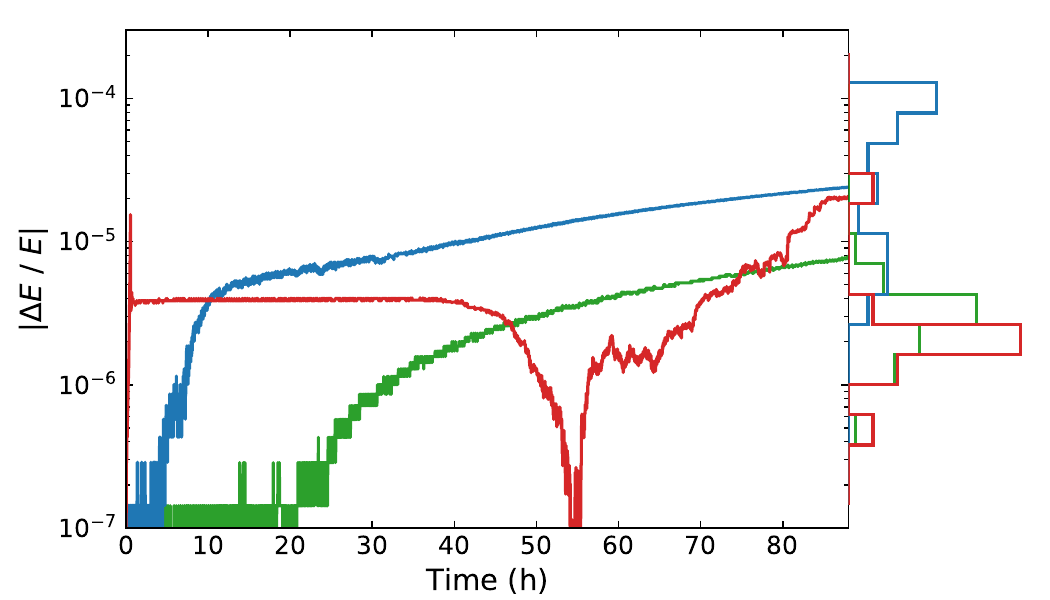}
\caption{Relative energy error ($|\Delta E / E|$) throughout the runtime for three example collisions : $\bimpact = 0.9$ at $\vimpact/\vesc \sim 0.78$ for \texttt{Hit \& Run} (green), $\bimpact = 0.14$ at $\vimpact/\vesc \sim 1$ representing a \texttt{Merger} (blue), and $\bimpact = 0.19$ at $\vimpact/\vesc \sim 5$ depicting a \texttt{Catastrophic} collision. The marginalized histograms represent the distribution of $|\Delta E / E|$ at simulation stopping time. 
}
\label{fig:energy_conservation}
\end{figure}

As shown in \autoref{fig:energy_conservation}, our simulations achieve good energy conservation. The 16th percentile, median, and 84th percentile values for the relative energy error $|\Delta E/ E|$ are $1.6\times 10^{-6}$, $3\times 10^{-6}$, and $2\times 10^{-5}$, respectively.

\begin{figure}[htb!]
\includegraphics[width=\columnwidth]{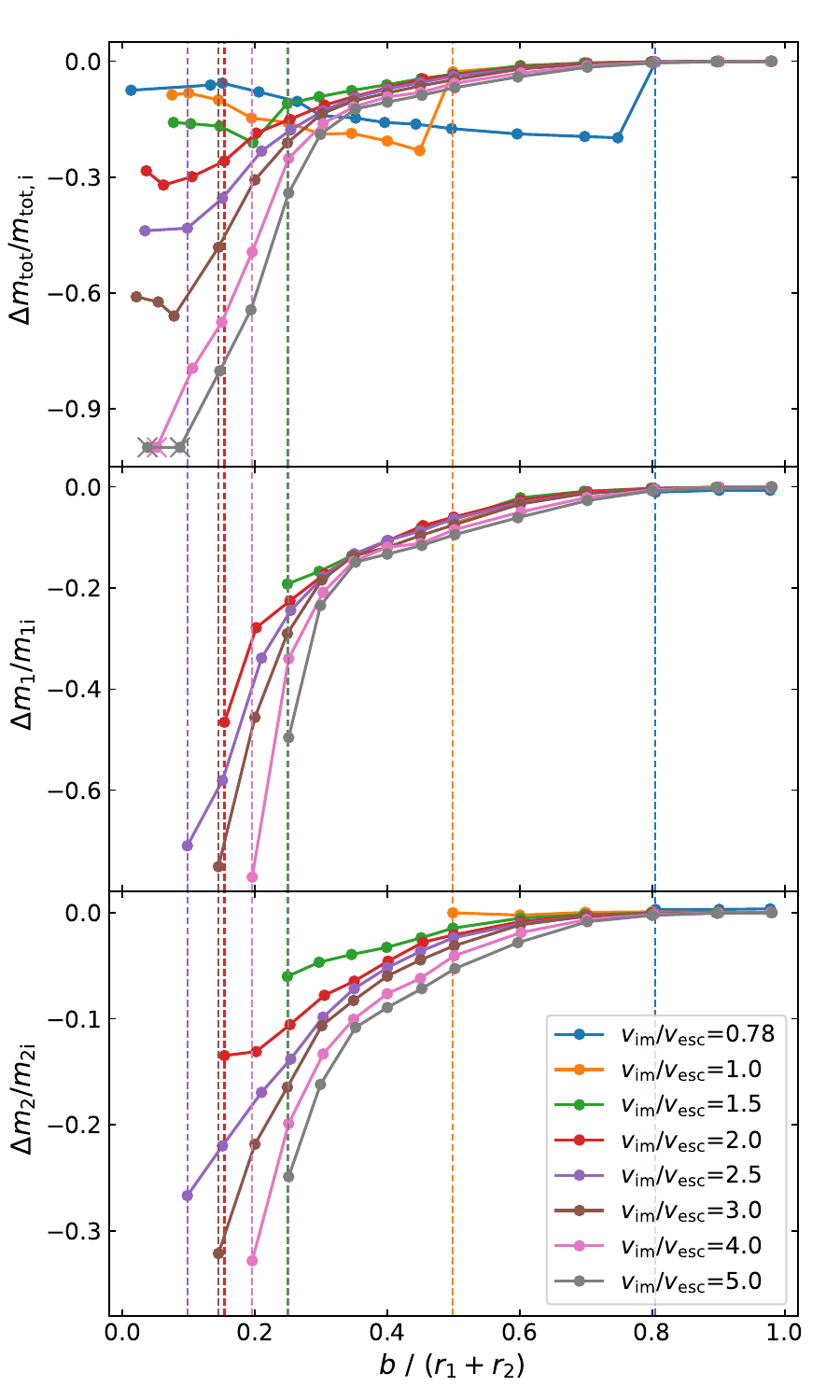}
\caption{Fractional change in planet masses, against impact parameter for different impact velocities. The vertical dashed lines indicate the lowest impact parameters at which both planets survive for each $\vimpact/\vesc$ (see legend). Collisions resulting in the destruction of both planets are marked by crosses.
}
\label{fig:dmp}
\end{figure}
%

\subsection{Hit \& Run Collisions}\label{subsec:results/Hit_Run}

In \texttt{Hit \& Run} collisions, both planets survive the first contact with sufficient momentum to overcome their mutual gravitational attraction and go on their separate orbits (top panel, \autoref{fig:snapshots}). There is no immediate re-encounter before the next possible orbit crossing. When the impact velocity ($\vimpact/\vesc \lesssim 1$) is low, the \texttt{Hit \& Run} collisions occur at higher impact parameters ($\bimpact \gtrsim 0.5$). During such encounters, the low-density H/He envelopes of the planets primarily interact, resulting in minimal mass loss predominantly from the gaseous envelopes. The fractional change in total planetary masses ($\dMptotbyMptot$, the subscript $\rm{i}$ denoting initial values), as well as the individual planetary masses ($\dMponebyMpone$, $\dMptwobyMptwo$), and the fractional change in the gaseous envelopes ($\dMgtotbyMgtot$, $\dMgonebyMgone$ and $\dMgtwobyMgtwo$) during the collisions with varying $\bimpact$ and $\vimpact$, are shown in \autoref{fig:dmp} and \autoref{fig:dmgas}, respectively. As the $\vimpact$ increases, the range of \texttt{Hit \& Run} encounters expands to include lower impact parameters. When $\vimpact$ exceeds $\vesc$, the mantles of the planets may interact and still have enough momentum to escape their mutual gravitational attraction (e.g., $\bimpact \sim 0.2$ with $\vimpact/\vesc \sim 2$). As expected, with even higher impact velocities, encounters featuring even smaller impact parameters can lead to a \texttt{Hit \& Run} collision and the mass loss increases as we increase $\vimpact/\vesc$ or decrease $\bimpact$ (see \autoref{fig:dmp}). During highly energetic collisions with low $\bimpact$, substantial portions of the mantle layers and gaseous envelopes of both planets can erode, with the less massive planet being particularly susceptible. In some extreme cases, e.g., at $\bimpact=0.14$ with $\vimpact/\vesc \sim 3$, the lower mass planet can lose $\gtrsim 80\%$ of its mantle, while the more massive planet still retains $\gtrsim 70\%$ of its mantle. For some specific \texttt{Hit \& Run} encounter (e.g., at $b \lesssim 0.35$ with $\vimpact/\vesc = 1.5$), the higher mass planet can even steal some of the mantle material from the lower mass planet (see \autoref{tab:all_sims}). Such energetic mantle-eroding collisions can potentially create the so-called `Super-Mercuries' \citep[e.g.,][]{2022_Reinhardt}. In general, the core of the planets remains relatively unaffected in \texttt{Hit \& Run} encounters with high impact parameters ($\bimpact \gtrsim 0.3$). However, for more energetic ($\vimpact/\vesc \gtrsim 2$) encounters with lower impact parameters ($\bimpact \lesssim 0.2$), the less massive planet can lose material from its core (see \autoref{tab:all_sims}). In some cases, the massive planet can even capture a small portion of the lower mass planet's core (e.g., at $b = 0.1$ with $\vimpact/\vesc = 2.5$) during the encounter.

\subsection{Mergers}\label{subsec:results/Merger}

As the impact parameter decreases, the amount of mass involved in the interaction during the encounter increases, making it progressively harder for the planets to overcome each other's gravitational pull. Consequently, this raises the probability of a \texttt{Merger}, where both the planets merge to form a single planet, more massive than either of the colliding planets. When the impact parameter and the impact velocity are sufficiently low, the planets collide nearly head-on and merge immediately, losing a small fraction of the total mass (e.g., $\dMptotbyMptot = -0.09$ for a collision at $\bimpact = 0.08$ with $\vimpact/\vesc \sim 1$, demonstrated in the middle panel of \autoref{fig:snapshots}). At greater impact velocities with comparable impact parameters, the degree of mass loss becomes more pronounced (e.g., $\dMptotbyMptot = -0.32$ for a collision at $\bimpact = 0.06$ with $\vimpact/\vesc \sim 2$). However, at higher impact parameters, the planets may not immediately merge during their initial close encounter. But, they also lack the necessary momentum to break free from each other's gravitational attraction, as in \texttt{Hit \& Run} encounters. After the first encounter, they return for one or more subsequent close encounters and eventually merge (e.g., the collision at $\bimpact = 0.4$ with $\vimpact/\vesc \sim 1$). This violent path to their eventual merger via multiple close approaches creates greater churning and results in a substantial mass loss (up to $23\%$ for $\vimpact/\vesc \lesssim 1.5$) in such events (see \autoref{fig:dmp} and \autoref{tab:all_sims}). Due to the repetitive nature of these encounters, it was necessary to extend these simulations beyond our typical stopping time, as detailed in \autoref{subsec:setup/sim_stopping}. Such collisions are often referred to as \texttt{Graze \& Merge} collisions \citep[e.g.,][]{2012_Genda, 2019_Cambioni, 2019_Emsenhuber_Asphaug}. Nevertheless, we treat these collisions as a subset of the \texttt{Mergers} due to the blurred boundary between these two.

\begin{figure}[htb!]
\includegraphics[width=\columnwidth]{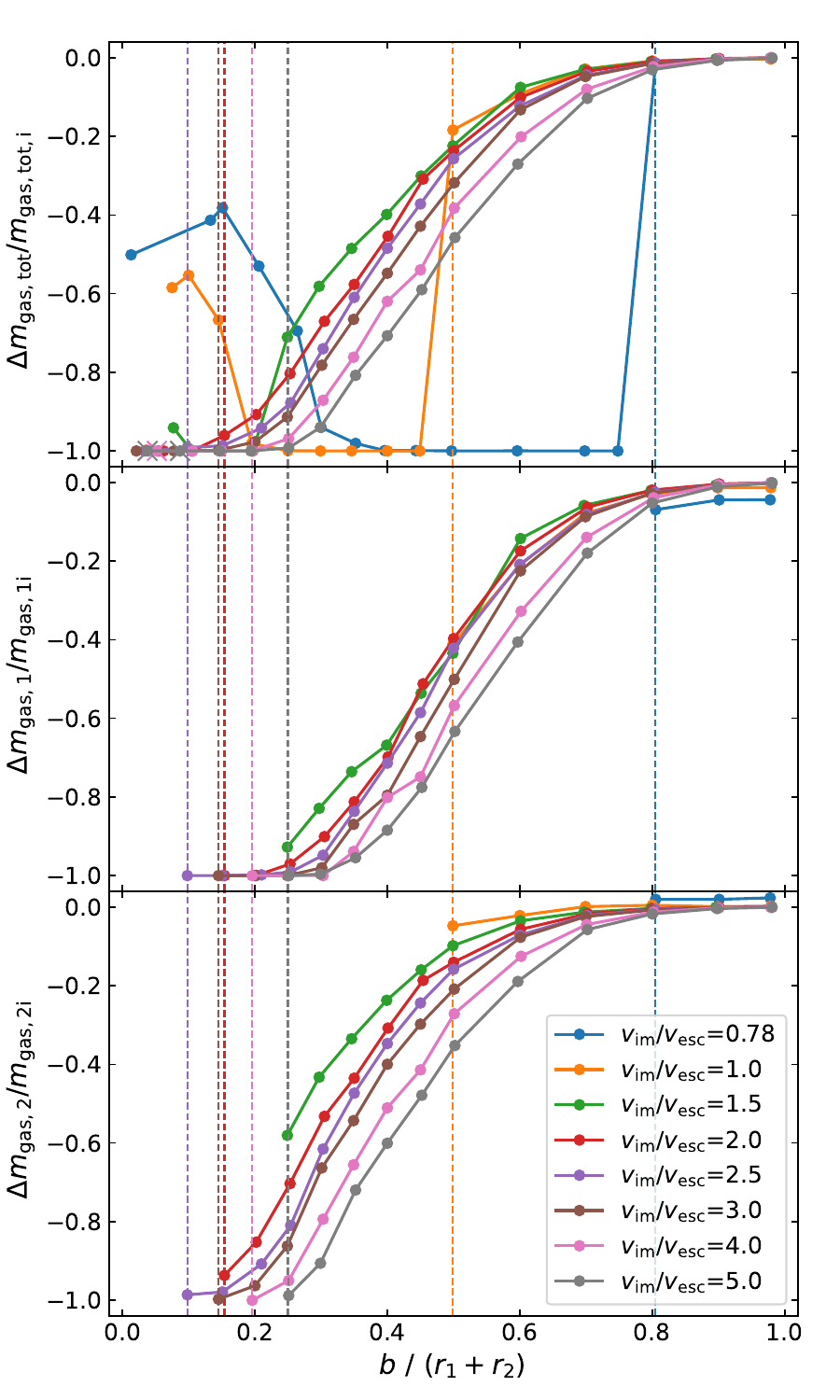}
\caption{Similar to \autoref{fig:dmp} but shows fractional change in the atmospheric mass of the planets with varying $\bimpact$ for different values of $\vimpact/\vesc$.
}
\label{fig:dmgas}
\end{figure}

Because of the increased mass loss in such collisions, the $\dMptotbyMptot$ versus $\bimpact$ curve shows a notable drop near the boundary between the \texttt{Hit \& Run} and \texttt{Merger} regimes (\autoref{fig:dmp}). For lower values of $\bimpact$, the collision smoothly transitions toward head-on \texttt{Merger}s, and the amount of mass lost during the encounters decreases. During \texttt{Merger} encounters with $\vimpact/\vesc \lesssim 2$, the collision remnant can retain the entire core material from the colliding planets, with the mass loss occurring solely from their mantles and gaseous envelopes. In such scenarios, it is possible for up to approximately $\sim30\%$ of the total mantle material to be stripped away along with the complete removal of their gaseous envelopes (e.g., at $\bimpact = 0.06$ with $\vimpact/\vesc \sim 2$). The fractional change in the total gas mass ($\dMgtotbyMgtot$) during these \texttt{Merger} events shows an interesting trend. For relatively larger $\bimpact$, particularly near the boundary between the \texttt{Hit \& Run} and \texttt{Merger} outcomes, where it takes repeated encounters for the planets to merge, the collision remnant cannot hold on to the atmosphere from the colliding bodies, resulting in a rocky collision product with no significant gaseous envelope. However, with lower impact parameters, the number of encounters needed to complete the merger reduces. Consequently, the collision remnant can retain substantial amounts of the atmospheres (up to $\sim 60\%$, e.g., at $\bimpact=0.15$ with $\vimpact/\vesc= 0.78$) from the parent bodies. The lower the impact velocity, the higher the amount of the atmosphere retained by the merger product. This shift and reversal in trends in both the $\dMptotbyMptot$ versus $\bimpact$ curve (\autoref{fig:dmp}) and the $\dMgtotbyMgtot$ versus $\bimpact$ curve (\autoref{fig:dmgas}) are particularly prominent when $\vimpact/\vesc \lesssim 1.5$.

\subsection{Catastrophic Collisions}\label{subsec:results/Catastrophic}

At high impact velocities ($\vimpact/\vesc \gtrsim 2.5$), collisions with low impact parameters ($\bimpact \lesssim 0.2$) can be highly destructive. Such high impact velocity collisions involve significant energy capable of ejecting substantial portions of the initially bound planetary mass beyond the gravitational influence of the collision remnant. In the aftermath of such collisions, if only one planetary body remains or none at all, and the total mass loss exceeds the initial mass of one of the colliding planets, signifying that the collision essentially destroyed one or both of them, we categorize the encounter as a \texttt{Catastrophic} collision
\footnote{Note that in some highly energetic collisions, the mentioned mass loss criteria can still be satisfied with both planets surviving the encounter (e.g., at $\bimpact=0.1$ with $\vimpact/\vesc \sim 2.5$). Nevertheless,  we treat them as \texttt{Hit \& Run} collisions since both the planets survive the encounter.}.
These collisions are also sometimes referred to as \texttt{Erosion} or \texttt{Disruption}. \citep[e.g.,][]{2012_Leinhardt, 2019_Cambioni}.
When a planetary body manages to survive such a highly destructive event, it can only retain a small portion of the combined mass from the colliding bodies. This results in significant loss of core and mantle materials and the complete depletion of the gaseous envelope that the parent bodies once had. In certain instances, the surviving body may consist of $\sim50\%$ ($\sim10\%$) of the combined initial core (mantle) material of the parent bodies (e.g., at $\bimpact=0.15$ with $\vimpact/\vesc \sim 5$). 
In some extreme scenarios, such as at $\bimpact \lesssim 0.05$ with $\vimpact/\vesc \sim 4$ (bottom panel, \autoref{fig:snapshots}) or at $\bimpact \lesssim 0.1$ with $\vimpact/\vesc \sim 5$, both parent planets disintegrate entirely, and the material that once was part of these planets starts orbiting the host star as debris.

Note that such exotic collisions between planetary bodies are exceptionally rare in typical simulations of dynamical instabilities that are expected to occur during the final stage of planet assembly; only one out of $1,094$ collisions in the ensemble \texttt{n8-e040-i024} of \citet{2023_Ghosh_PPScattering} falls under this category (private communication).

\subsection{Boundaries Between Different Outcomes as a Function of $\bimpact$ and $\vimpact/\vesc$}\label{subsec:results/boundaries}

\begin{figure}[htb!]
\includegraphics[width=\columnwidth]{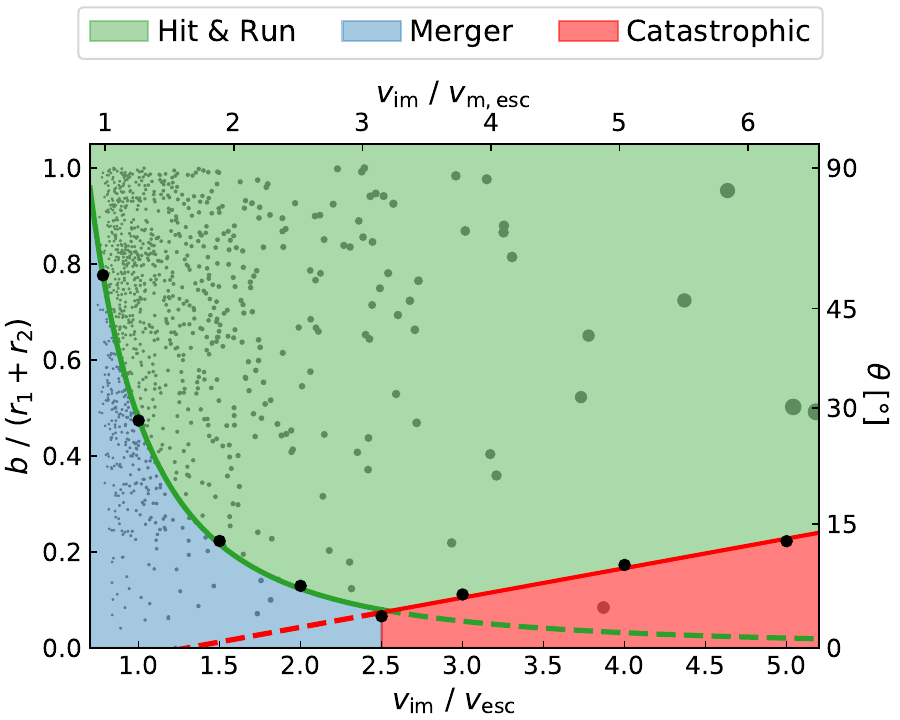}
\caption{Summary of overall outcomes in $\bimpact - \vimpact$ space. The black dots mark the $\bcrit$ values from our simulation grid. The boundary separating \texttt{Hit \& Run} collisions with other outcomes, follows $\bcrit/\bcritvesc \approx (\vimpact/\vesc)^{-2}$, for $\vimpact/\vesc \lesssim 2.5$ and $\bcrit$ increases linearly ($\bcrit/\bcritvesctwo \approx (\vimpact/\vesc) - 1.5$) at higher impact velocities. Similar to \autoref{fig:grid}, the grey dots represent the collisions produced in typical planet-planet scattering experiments (ensemble \texttt{n8-e040-i024} of \citet{2023_Ghosh_PPScattering}).
}
\label{fig:outcomes}
\end{figure}

In this section, we closely examine the boundaries between different outcomes in the $\bimpact-\vimpact$ space. The three major regimes in the $\bimpact-\vimpact/\vesc$ space are summarized in \autoref{fig:outcomes}. We first examine the critical value, $\bcrit$ (prime indicating normalization by $\Rpone + \Rptwo$) of $\bimpact$ separating the \texttt{Hit \& Run} encounters with others for the different values of $\vimpact/\vesc$. In our discrete grid, we estimate $\bcrit$ as the midpoint between the adjacent two grid points where the outcome transitions for a given $\vimpact/\vesc$. As expected, $\bcrit$ depends on $\vimpact$. In general, lower $\vimpact$ leads to higher $\bcrit$. For example, in the collisions with $\vimpact/\vesc= 2$, we find $\bcrit=0.13$, that is, all encounters with $\bimpact < 0.13$ lead to the merger of the two planets. On the other hand, for $\vimpact/\vesc= 0.78$, $\bcrit=0.78$.

For $\vimpact/\vesc \lesssim 2.5$, we find that the dependency of $\bcrit$ on $\vimpact/\vesc$ approximately follows $\bcrit = \bcritvesc \times (\vimpact/\vesc)^{-2}$, where $\bcritvesc = 0.474 \pm 0.025$ is the value of $\bcrit$ at $\vimpact/\vesc = 1$. The errors indicate the spacing between our discrete grid points near $\bcrit$. In other words $\bcrit$ is inversely proportional to the impact energy ($\Eimpact = \frac{1}{2}\mu \vimpact^{2}$, where $\mu = \frac{\Mpone \Mptwo}{\Mpone + \Mptwo}$, the reduced mass of the two planet system). This dependency arises since, at lower impact speeds or equivalently lower energies, the planets do not have enough energy to escape their mutual gravitational potential well, even at higher impact parameters, ultimately resulting in \texttt{Mergers}. Conversely, as the impact energy increases, the planets can penetrate more deeply into their gravitational potential well and still escape, resulting in lower $\bcrit$ for higher $\vimpact$.

Our results are qualitatively similar to the findings of previous studies. \citet{2012_Leinhardt, 2012_Stewart_Leinhardt} found that for gravity-dominated planetesimals, \texttt{Hit \& Run} collisions occur when the bulk of the impactor misses the target, that is $\bimpact > \Rptwo / (\Rpone + \Rptwo)$, assuming Planet-2 as the target and Planet-1 as the impactor. Additionally, for $\bimpact > \Rptwo / (\Rpone + \Rptwo)$ the boundary between the \texttt{Hit \& Run} and the \texttt{Graze \& Merge} scenarios is qualitatively similar to our $\bcrit \propto \vesc^{-2}$ boundary \citep[see Figure~5 of][]{2012_Stewart_Leinhardt}. However, for more massive density-stratified bodies, \texttt{Hit \& Run} collisions can occur at much lower impact parameters \citep{2020_Gabriel}. The planets considered in our study are massive differentiated sub-Neptunes where most of the mass is densely concentrated towards the center, and a very low-density H/He envelope comprises most of its volume. Consequently, the boundary between the \texttt{Hit \& Run} and \texttt{Merger} regimes occur at much lower values of $\bimpact$.
Recently, \citet{2022_Denman} considered collisions involving a massive planet with significant H/He atmospheres and atmosphere-less projectiles. They have found that the \texttt{Hit \& Run} collisions occurs at $\vimpact > \sqrt{2 G (\Mpone + \Mptwo)/ b}$. Although this is qualitatively similar to our finding of $\bcrit \propto \vesc^{-2}$, the values of $\bcrit$ in our study are lower than their estimation. For example, their finding indicates that at $\vimpact \sim \vesc$, \texttt{Hit \& Run} collisions would occur at $\bimpact > 0.63$, higher than the $\bcritvesc$ we found. This discrepancy is likely because both planets in our study had substantial H/He atmospheres, whereas \citet{2022_Denman} considered atmosphere-less projectiles.
Interestingly, for the planets considered in our study, we find $\bcritvesc \approx 1.2 \times (\Rm_{\rm{1}} + \Rm_{\rm{2}})$. It would be interesting to investigate whether this relationship holds for planets with different structures and initial properties in future studies.

Around $\vimpact/\vesc \approx 2.5$, $\bcrit$ decreases significantly and reaches $\bcritvesctwo=0.07$, equivalent to $\approx (\Rm_{\rm{1}} + \Rm_{\rm{2}})/6$. In such high $\vimpact$ collisions at low $\bimpact$ ($\lesssim \bcrit$), the outcome is typically a \texttt{Catastrophic} collision. Beyond $\vimpact/\vesc \approx 2.5$, $\bcrit$ no longer decreases with increasing $\vimpact$. Instead $\bcrit$ exhibits a linear increase with $\vimpact$: $\bcrit/\bcritvesctwo \approx (\vimpact/\vesc) - 1.5$. In this regime, $\bcrit$ is so low that significant portions of the mantle and the core are involved in encounters with $\bimpact \approx \bcrit$. At lower energies with $\bimpact$ just above $\bcrit$, despite significant erosion, both planets can still survive the encounter, resulting in a \texttt{Hit \& Run} collision (e.g., at $\bimpact=0.14$ with $\vimpact/\vesc\sim 3$). However, at the same $b$ at a higher impact velocity (e.g., at $\bimpact=0.15$ with $\vimpact/\vesc\sim 4$), the increased energy of the impact causes the less massive planet to disintegrate entirely, ultimately resulting in only one survivor. Consequently, $\bcrit$ increases with $\vimpact/\vesc$.

The parameter space associated with the \texttt{Catastrophic} regime qualitatively resembles the \texttt{Erosion} and \texttt{Disruption} regime observed in terrestrial protoplanet collisions \citep[see][]{2019_Cambioni, 2020_Emsenhuber, 2023_Gabriel_annurev}. However, due to the different structures of the planets considered in our study, the transition from \texttt{Hit \& Run} regime occurs at much lower impact parameters. For example, at $\vimpact/\vesc \sim 3$, the transition occurs at $\bimpact \sim 0.1$, whereas for terrestrial protoplanets it occurs around $\bimpact \sim 0.35$ \citep{2020_Emsenhuber}.

\section{Summary And Discussions}\label{sec:summary}

In this study, we present our numerical framework to study sub-Neptune collisions using a combination of $N$-body, planet structure, and SPH simulations. We have applied our setup to conduct $123$ hydrodynamical simulations of collisions between two specific sub-Neptune planets for various impact parameters and impact velocities expected during dynamical instabilities. Our key results are summarized below.
\begin{itemize}

  \item There are three different outcomes, namely a) \texttt{Hit \& Run} in which both planets survive the grazing encounter, b) \texttt{Mergers}, forming a single planet, more massive than the individual colliding planets, and c) \texttt{Catastrophic} collisions, resulting in the effective destruction of one or both planets.

  \item \texttt{Hit \& Run} encounters are the most common, spanning the most substantial region in the $\bimpact-\vimpact$ space (\autoref{fig:outcomes}), among the three. Such collisions occur at relatively larger impact parameters at lower impact velocities, where only the low-density gaseous envelopes interact. As impact velocities increase, these encounters extend to smaller $\bimpact$. The higher the impact velocity or lower the impact parameter, the greater the amount of mass lost during the encounter.

  \item At lower impact parameters, the planets cannot overcome their mutual gravitational pull, resulting in a \texttt{Merger}, where they form a single massive planet, losing a small fraction of the total mass of the colliding planets. At relatively higher impact parameters, the planets may survive the initial encounter. However, without sufficient energy to overcome their gravitational attraction, they would return immediately for one or more close encounters and eventually merge with more substantial mass loss.

  \item At large impact velocities ($\vimpact/\vesc \gtrsim 2.5$), the low-$\bimpact$ collisions can be highly destructive (red shaded region, \autoref{fig:outcomes}), effectively disintegrating one or, in some extreme cases, both planets. These collisions are referred to as \texttt{Catastrophic} collisions.

  \item The critical impact parameter ($\bcrit$) separating the \texttt{Hit \& Run} and \texttt{Merger} encounters, depends on $\vimpact$ or equivalently the impact energy $\Eimpact$. We find that for $\vimpact/\vesc \lesssim 2.5$, $\bcrit\propto(\vimpact/\vesc)^{-2}\sim\Eimpact^{-1}$. In other words, as the planets get closer, they require more energy to overcome their gravitational attraction. For $\vimpact/\vesc \gtrsim 2.5$, $\bcrit$ separates the \texttt{Hit \& Run} and \texttt{Catastrophic} encounters. In this regime, $\bcrit$ increases linearly with  ($\vimpact/\vesc$) as the increased impact energy disintegrates the lower mass planet more efficiently.
  
  \item Interestingly, we find that the majority of the collisions ($\approx76\%$) that were treated as perfect mergers in \cite{2023_Ghosh_PPScattering} may fall within the \texttt{Hit \& Run} regime if the collisions involve sub-Neptunes with substantial gaseous envelopes. Even those falling under the \texttt{Mergers} regime, the remnants can lose a significant amount of mass and atmosphere and hence can not be treated as perfect mergers.

\end{itemize}

A more detailed future study with a multi-dimensional grid with the initial planet masses and atmospheric fractions as input parameters would shed more light on the outcomes of the sub-Neptune collisions. Such a study would provide insights into the amount of mass loss and atmospheric depletion in such encounters for a wider range in parameter space, offering valuable scaling laws applicable to sub-Neptunes. The development of an analytic prescription based on these results could be of great use to enable us treat collisions more realistically in $N$-body modeling of planetary systems. It would be particularly interesting to assess the impact of accurate collision modeling on the outcomes of $N$-body dynamical evolution studies of exoplanets, such as \cite{2023_Ghosh_PPScattering}. This work is the first step towards this goal.


\begin{acknowledgments}
We thank the anonymous referee for insightful comments and helpful suggestions to improve the manuscript. TG acknowledges support from TIFR's graduate fellowship. SC acknowledges support from the Department of Atomic Energy, Government of India, under project no. 12-R\&D-TFR-5.02-0200 and RTI 4002. We would like to thank Jason Hwang for useful discussions while setting up the SPH simulations. 
This work made use of the SPLASH visualization software \citep{price_2007}.
\end{acknowledgments}


\software{\texttt{Rebound} \citep{2012_REB}, \texttt{WoMa} \citep{2020_WoMa}, \texttt{StarSmasher} \citep{2010_Gaburov, 2018_Gaburov_Code}, \texttt{NumPy} \citep{2020_numpy}, \texttt{SciPy} \citep{2020SciPy-NMeth}, \texttt{pandas} \citep{pandas, mckinney-proc-scipy-2010}, \texttt{scikit-learn} \citep{scikit-learn}, \texttt{matplotlib} \citep{matplotlib_Hunter:2007}, \texttt{SPLASH} \citep{price_2007}}


\centerwidetable
\startlongtable
\begin{deluxetable*}{cccccccccc}
\tabletypesize{\footnotesize}
\tablecaption{Summarized results of all the simulations conducted in this study.\label{tab:all_sims}}

\tablehead{
    \colhead{Number} &
    \colhead{$v_{\rm{im}}/v_{\rm{esc}}$} &
    \colhead{$b\  /\ (r_{1} + r_{2})$} &
    \colhead{$t_{\rm{stop}}/\rm{hour}$} &
    \colhead{Outcome} &
    \colhead{$m$} &
    \colhead{$\Delta m / m_{\rm{i}}$} &
    \colhead{$\Delta m_{\rm{core}} / m_{\rm{core,i}}$} &
    \colhead{$\Delta m_{\rm{mantle}} / m_{\rm{mantle,i}}$} &
    \colhead{$\Delta m_{\rm{gas}} / m_{\rm{gas,i}}$}
    }

\startdata
     1 &                     $0.78$ &                   $0.01$ &                    $44.3$ &      MG &        $11.14$ &                  $-0.08$ &                                    $0.00$ &                                        $0.00$ &                                 $-0.50$ \\
     2 &                     $0.77$ &                   $0.13$ &                    $44.3$ &      MG &        $11.30$ &                  $-0.06$ &                                    $0.00$ &                                        $0.00$ &                                 $-0.41$ \\
     3 &                     $0.78$ &                   $0.15$ &                    $44.3$ &      MG &        $11.36$ &                  $-0.06$ &                                    $0.00$ &                                        $0.00$ &                                 $-0.38$ \\
     4 &                     $0.76$ &                   $0.21$ &                   $132.8$ &      MG &        $11.08$ &                  $-0.08$ &                                    $0.00$ &                                        $0.00$ &                                 $-0.53$ \\
     5 &                     $0.76$ &                   $0.26$ &                   $177.1$ &      MG &        $10.79$ &                  $-0.10$ &                                    $0.00$ &                                        $0.00$ &                                 $-0.69$ \\
     6 &                     $0.78$ &                   $0.30$ &                   $177.1$ &      MG &        $10.35$ &                  $-0.14$ &                                    $0.00$ &                                        $0.00$ &                                 $-0.94$ \\
     7 &                     $0.77$ &                   $0.35$ &                   $177.1$ &      MG &        $10.27$ &                  $-0.15$ &                                    $0.00$ &                                        $0.00$ &                                 $-0.98$ \\
     8 &                     $0.77$ &                   $0.40$ &                   $265.6$ &      MG &        $10.13$ &                  $-0.16$ &                                    $0.00$ &                                       $-0.01$ &                                 $-1.00$ \\
     9 &                     $0.77$ &                   $0.44$ &                   $265.6$ &      MG &        $10.08$ &                  $-0.16$ &                                    $0.00$ &                                       $-0.02$ &                                 $-1.00$ \\
    10 &                     $0.76$ &                   $0.50$ &                   $332.0$ &      MG &         $9.94$ &                  $-0.17$ &                                    $0.00$ &                                       $-0.04$ &                                 $-1.00$ \\
    11 &                     $0.78$ &                   $0.60$ &                   $332.0$ &      MG &         $9.78$ &                  $-0.19$ &                                    $0.00$ &                                       $-0.06$ &                                 $-1.00$ \\
    12 &                     $0.78$ &                   $0.70$ &                   $332.0$ &      MG &         $9.70$ &                  $-0.19$ &                                    $0.00$ &                                       $-0.08$ &                                 $-1.00$ \\
    13 &                     $0.77$ &                   $0.75$ &                   $332.0$ &      MG &         $9.66$ &                  $-0.20$ &                                    $0.00$ &                                       $-0.08$ &                                 $-1.00$ \\
    14 &                     $0.78$ &                   $0.80$ &                    $44.3$ &      HR &  $4.46,\ 7.56$ &           $-0.01,\ 0.00$ &                             $0.00,\ 0.00$ &                                 $0.00,\ 0.00$ &                          $-0.07,\ 0.02$ \\
    15 &                     $0.78$ &                   $0.90$ &                    $88.5$ &      HR &  $4.47,\ 7.56$ &           $-0.01,\ 0.00$ &                             $0.00,\ 0.00$ &                                 $0.00,\ 0.00$ &                          $-0.04,\ 0.02$ \\
    16 &                     $0.78$ &                   $0.98$ &                    $44.3$ &      HR &  $4.47,\ 7.57$ &           $-0.01,\ 0.00$ &                             $0.00,\ 0.00$ &                                 $0.00,\ 0.00$ &                          $-0.04,\ 0.02$ \\
    17 &                     $0.99$ &                   $0.08$ &                    $44.3$ &      MG &        $10.99$ &                  $-0.09$ &                                    $0.00$ &                                        $0.00$ &                                 $-0.58$ \\
    18 &                     $0.98$ &                   $0.10$ &                    $44.3$ &      MG &        $11.05$ &                  $-0.08$ &                                    $0.00$ &                                        $0.00$ &                                 $-0.55$ \\
    19 &                     $1.00$ &                   $0.14$ &                    $88.5$ &      MG &        $10.84$ &                  $-0.10$ &                                    $0.00$ &                                       $-0.00$ &                                 $-0.67$ \\
    20 &                     $1.04$ &                   $0.20$ &                   $221.3$ &      MG &        $10.27$ &                  $-0.15$ &                                    $0.00$ &                                        $0.00$ &                                 $-0.98$ \\
    21 &                     $0.98$ &                   $0.25$ &                   $265.6$ &      MG &        $10.13$ &                  $-0.16$ &                                    $0.00$ &                                       $-0.01$ &                                 $-1.00$ \\
    22 &                     $1.02$ &                   $0.30$ &                   $332.0$ &      MG &         $9.77$ &                  $-0.19$ &                                    $0.00$ &                                       $-0.06$ &                                 $-1.00$ \\
    23 &                     $1.01$ &                   $0.35$ &                   $332.0$ &      MG &         $9.80$ &                  $-0.19$ &                                    $0.00$ &                                       $-0.06$ &                                 $-1.00$ \\
    24 &                     $0.97$ &                   $0.40$ &                   $332.0$ &      MG &         $9.56$ &                  $-0.21$ &                                    $0.00$ &                                       $-0.09$ &                                 $-1.00$ \\
    25 &                     $1.00$ &                   $0.45$ &                   $332.0$ &      MG &         $9.26$ &                  $-0.23$ &                                   $-0.00$ &                                       $-0.14$ &                                 $-1.00$ \\
    26 &                     $1.00$ &                   $0.50$ &                    $44.3$ &      HR &  $4.18,\ 7.54$ &          $-0.07,\ -0.00$ &                             $0.00,\ 0.00$ &                                $-0.02,\ 0.01$ &                         $-0.41,\ -0.05$ \\
    27 &                     $1.04$ &                   $0.60$ &                    $44.3$ &      HR &  $4.35,\ 7.52$ &          $-0.03,\ -0.00$ &                             $0.00,\ 0.00$ &                                $-0.00,\ 0.00$ &                         $-0.21,\ -0.02$ \\
    28 &                     $1.03$ &                   $0.70$ &                    $44.3$ &      HR &  $4.45,\ 7.54$ &           $-0.01,\ 0.00$ &                             $0.00,\ 0.00$ &                                 $0.00,\ 0.00$ &                          $-0.08,\ 0.00$ \\
    29 &                     $1.02$ &                   $0.80$ &                    $44.3$ &      HR &  $4.48,\ 7.54$ &           $-0.00,\ 0.00$ &                             $0.00,\ 0.00$ &                                 $0.00,\ 0.00$ &                          $-0.03,\ 0.00$ \\
    30 &                     $1.04$ &                   $0.90$ &                    $44.3$ &      HR &  $4.49,\ 7.54$ &           $-0.00,\ 0.00$ &                             $0.00,\ 0.00$ &                                 $0.00,\ 0.00$ &                          $-0.01,\ 0.00$ \\
    31 &                     $0.95$ &                   $0.98$ &                    $44.3$ &      HR &  $4.49,\ 7.54$ &           $-0.00,\ 0.00$ &                             $0.00,\ 0.00$ &                                 $0.00,\ 0.00$ &                          $-0.01,\ 0.00$ \\
    32 &                     $1.48$ &                   $0.08$ &                    $88.5$ &      MG &        $10.13$ &                  $-0.16$ &                                    $0.00$ &                                       $-0.03$ &                                 $-0.94$ \\
    33 &                     $1.46$ &                   $0.10$ &                   $177.1$ &      MG &        $10.10$ &                  $-0.16$ &                                    $0.00$ &                                       $-0.02$ &                                 $-1.00$ \\
    34 &                     $1.47$ &                   $0.15$ &                   $177.1$ &      MG &        $10.02$ &                  $-0.17$ &                                    $0.00$ &                                       $-0.03$ &                                 $-1.00$ \\
    35 &                     $1.53$ &                   $0.20$ &                   $265.6$ &      MG &         $9.50$ &                  $-0.21$ &                                    $0.00$ &                                       $-0.10$ &                                 $-1.00$ \\
    36 &                     $1.51$ &                   $0.25$ &                    $44.3$ &      HR &  $3.64,\ 7.09$ &          $-0.19,\ -0.06$ &                            $-0.02,\ 0.01$ &                                $-0.08,\ 0.04$ &                         $-0.93,\ -0.58$ \\
    37 &                     $1.49$ &                   $0.30$ &                    $44.3$ &      HR &  $3.75,\ 7.19$ &          $-0.17,\ -0.05$ &                             $0.00,\ 0.00$ &                                $-0.07,\ 0.03$ &                         $-0.83,\ -0.43$ \\
    38 &                     $1.46$ &                   $0.35$ &                    $44.3$ &      HR &  $3.89,\ 7.24$ &          $-0.14,\ -0.04$ &                             $0.00,\ 0.00$ &                                $-0.04,\ 0.02$ &                         $-0.74,\ -0.33$ \\
    39 &                     $1.50$ &                   $0.40$ &                    $44.3$ &      HR &  $4.02,\ 7.29$ &          $-0.11,\ -0.03$ &                             $0.00,\ 0.00$ &                                $-0.01,\ 0.00$ &                         $-0.67,\ -0.24$ \\
    40 &                     $1.54$ &                   $0.45$ &                    $44.3$ &      HR &  $4.14,\ 7.36$ &          $-0.08,\ -0.02$ &                             $0.00,\ 0.00$ &                                 $0.00,\ 0.00$ &                         $-0.54,\ -0.16$ \\
    41 &                     $1.48$ &                   $0.50$ &                    $44.3$ &      HR &  $4.21,\ 7.43$ &          $-0.07,\ -0.01$ &                             $0.00,\ 0.00$ &                                 $0.00,\ 0.00$ &                         $-0.43,\ -0.10$ \\
    42 &                     $1.50$ &                   $0.60$ &                    $44.3$ &      HR &  $4.41,\ 7.50$ &          $-0.02,\ -0.01$ &                             $0.00,\ 0.00$ &                                 $0.00,\ 0.00$ &                         $-0.14,\ -0.04$ \\
    43 &                     $1.54$ &                   $0.70$ &                    $44.3$ &      HR &  $4.46,\ 7.52$ &          $-0.01,\ -0.00$ &                             $0.00,\ 0.00$ &                                 $0.00,\ 0.00$ &                         $-0.06,\ -0.01$ \\
    44 &                     $1.53$ &                   $0.80$ &                    $44.3$ &      HR &  $4.49,\ 7.53$ &          $-0.00,\ -0.00$ &                             $0.00,\ 0.00$ &                                 $0.00,\ 0.00$ &                         $-0.02,\ -0.00$ \\
    45 &                     $1.49$ &                   $0.90$ &                    $44.3$ &      HR &  $4.50,\ 7.54$ &          $-0.00,\ -0.00$ &                             $0.00,\ 0.00$ &                                 $0.00,\ 0.00$ &                         $-0.01,\ -0.00$ \\
    46 &                     $1.49$ &                   $0.98$ &                    $44.3$ &      HR &  $4.50,\ 7.54$ &          $-0.00,\ -0.00$ &                             $0.00,\ 0.00$ &                                 $0.00,\ 0.00$ &                         $-0.00,\ -0.00$ \\
    47 &                     $1.99$ &                   $0.04$ &                    $88.5$ &      MG &         $8.62$ &                  $-0.28$ &                                    $0.00$ &                                       $-0.23$ &                                 $-1.00$ \\
    48 &                     $2.04$ &                   $0.06$ &                    $88.5$ &      MG &         $8.18$ &                  $-0.32$ &                                   $-0.00$ &                                       $-0.29$ &                                 $-1.00$ \\
    49 &                     $2.02$ &                   $0.10$ &                    $88.5$ &      MG &         $8.44$ &                  $-0.30$ &                                    $0.00$ &                                       $-0.25$ &                                 $-1.00$ \\
    50 &                     $1.98$ &                   $0.15$ &                    $44.3$ &      HR &  $2.41,\ 6.52$ &          $-0.47,\ -0.13$ &                            $-0.10,\ 0.05$ &                               $-0.49,\ -0.01$ &                         $-1.00,\ -0.94$ \\
    51 &                     $2.01$ &                   $0.20$ &                    $44.3$ &      HR &  $3.25,\ 6.55$ &          $-0.28,\ -0.13$ &                            $-0.03,\ 0.01$ &                               $-0.20,\ -0.01$ &                         $-1.00,\ -0.85$ \\
    52 &                     $1.97$ &                   $0.25$ &                    $44.3$ &      HR &  $3.49,\ 6.74$ &          $-0.23,\ -0.11$ &                            $-0.01,\ 0.00$ &                               $-0.13,\ -0.00$ &                         $-0.97,\ -0.70$ \\
    53 &                     $1.97$ &                   $0.30$ &                    $44.3$ &      HR &  $3.73,\ 6.95$ &          $-0.17,\ -0.08$ &                             $0.00,\ 0.00$ &                                $-0.06,\ 0.00$ &                         $-0.90,\ -0.53$ \\
    54 &                     $2.04$ &                   $0.35$ &                    $44.3$ &      HR &  $3.89,\ 7.05$ &          $-0.14,\ -0.06$ &                             $0.00,\ 0.00$ &                                $-0.02,\ 0.00$ &                         $-0.81,\ -0.44$ \\
    55 &                     $1.99$ &                   $0.40$ &                    $44.3$ &      HR &  $4.02,\ 7.19$ &          $-0.11,\ -0.05$ &                             $0.00,\ 0.00$ &                                $-0.00,\ 0.00$ &                         $-0.70,\ -0.31$ \\
    56 &                     $2.02$ &                   $0.45$ &                    $44.3$ &      HR &  $4.16,\ 7.33$ &          $-0.08,\ -0.03$ &                             $0.00,\ 0.00$ &                                 $0.00,\ 0.00$ &                         $-0.51,\ -0.19$ \\
    57 &                     $2.00$ &                   $0.50$ &                    $44.3$ &      HR &  $4.23,\ 7.38$ &          $-0.06,\ -0.02$ &                             $0.00,\ 0.00$ &                                 $0.00,\ 0.00$ &                         $-0.40,\ -0.14$ \\
    58 &                     $2.02$ &                   $0.60$ &                    $44.3$ &      HR &  $4.39,\ 7.47$ &          $-0.03,\ -0.01$ &                             $0.00,\ 0.00$ &                                 $0.00,\ 0.00$ &                         $-0.17,\ -0.06$ \\
    59 &                     $2.04$ &                   $0.70$ &                    $44.3$ &      HR &  $4.46,\ 7.52$ &          $-0.01,\ -0.00$ &                             $0.00,\ 0.00$ &                                 $0.00,\ 0.00$ &                         $-0.06,\ -0.02$ \\
    60 &                     $1.96$ &                   $0.80$ &                    $44.3$ &      HR &  $4.49,\ 7.53$ &          $-0.00,\ -0.00$ &                             $0.00,\ 0.00$ &                                 $0.00,\ 0.00$ &                         $-0.02,\ -0.00$ \\
    61 &                     $2.00$ &                   $0.90$ &                    $44.3$ &      HR &  $4.50,\ 7.54$ &          $-0.00,\ -0.00$ &                             $0.00,\ 0.00$ &                                 $0.00,\ 0.00$ &                         $-0.00,\ -0.00$ \\
    62 &                     $2.04$ &                   $0.98$ &                    $44.3$ &      HR &  $4.50,\ 7.54$ &          $-0.00,\ -0.00$ &                             $0.00,\ 0.00$ &                                 $0.00,\ 0.00$ &                         $-0.00,\ -0.00$ \\
    63 &                     $2.46$ &                   $0.03$ &                    $44.3$ &      CC &         $6.76$ &                  $-0.44$ &                                   $-0.04$ &                                       $-0.47$ &                                 $-1.00$ \\
    64 &                     $2.54$ &                   $0.10$ &                    $44.3$ &      HR &  $1.31,\ 5.53$ &          $-0.71,\ -0.27$ &                            $-0.39,\ 0.12$ &                               $-0.77,\ -0.25$ &                         $-1.00,\ -0.99$ \\
    65 &                     $2.52$ &                   $0.15$ &                    $44.3$ &      HR &  $1.89,\ 5.88$ &          $-0.58,\ -0.22$ &                            $-0.14,\ 0.02$ &                               $-0.66,\ -0.13$ &                         $-1.00,\ -0.98$ \\
    66 &                     $2.50$ &                   $0.21$ &                    $44.3$ &      HR &  $2.98,\ 6.26$ &          $-0.34,\ -0.17$ &                            $-0.04,\ 0.00$ &                               $-0.30,\ -0.06$ &                         $-1.00,\ -0.91$ \\
    67 &                     $2.52$ &                   $0.25$ &                    $44.3$ &      HR &  $3.40,\ 6.50$ &          $-0.24,\ -0.14$ &                             $0.00,\ 0.00$ &                               $-0.16,\ -0.03$ &                         $-0.99,\ -0.81$ \\
    68 &                     $2.51$ &                   $0.30$ &                    $44.3$ &      HR &  $3.70,\ 6.80$ &          $-0.18,\ -0.10$ &                             $0.00,\ 0.00$ &                               $-0.06,\ -0.01$ &                         $-0.95,\ -0.62$ \\
    69 &                     $2.51$ &                   $0.35$ &                    $44.3$ &      HR &  $3.91,\ 7.00$ &          $-0.13,\ -0.07$ &                             $0.00,\ 0.00$ &                               $-0.01,\ -0.00$ &                         $-0.84,\ -0.47$ \\
    70 &                     $2.49$ &                   $0.40$ &                    $44.3$ &      HR &  $4.02,\ 7.15$ &          $-0.11,\ -0.05$ &                             $0.00,\ 0.00$ &                                 $0.00,\ 0.00$ &                         $-0.71,\ -0.35$ \\
    71 &                     $2.52$ &                   $0.45$ &                    $44.3$ &      HR &  $4.11,\ 7.26$ &          $-0.09,\ -0.04$ &                             $0.00,\ 0.00$ &                                 $0.00,\ 0.00$ &                         $-0.59,\ -0.24$ \\
    72 &                     $2.52$ &                   $0.50$ &                    $44.3$ &      HR &  $4.22,\ 7.36$ &          $-0.06,\ -0.02$ &                             $0.00,\ 0.00$ &                                 $0.00,\ 0.00$ &                         $-0.42,\ -0.16$ \\
    73 &                     $2.50$ &                   $0.60$ &                    $44.3$ &      HR &  $4.36,\ 7.46$ &          $-0.03,\ -0.01$ &                             $0.00,\ 0.00$ &                                 $0.00,\ 0.00$ &                         $-0.21,\ -0.07$ \\
    74 &                     $2.47$ &                   $0.70$ &                    $44.3$ &      HR &  $4.45,\ 7.51$ &          $-0.01,\ -0.00$ &                             $0.00,\ 0.00$ &                                 $0.00,\ 0.00$ &                         $-0.08,\ -0.02$ \\
    75 &                     $2.47$ &                   $0.80$ &                    $44.3$ &      HR &  $4.49,\ 7.53$ &          $-0.00,\ -0.00$ &                             $0.00,\ 0.00$ &                                 $0.00,\ 0.00$ &                         $-0.02,\ -0.01$ \\
    76 &                     $2.52$ &                   $0.90$ &                    $44.3$ &      HR &  $4.50,\ 7.54$ &          $-0.00,\ -0.00$ &                             $0.00,\ 0.00$ &                                 $0.00,\ 0.00$ &                         $-0.01,\ -0.00$ \\
    77 &                     $2.55$ &                   $0.98$ &                    $44.3$ &      HR &  $4.50,\ 7.54$ &          $-0.00,\ -0.00$ &                             $0.00,\ 0.00$ &                                 $0.00,\ 0.00$ &                         $-0.00,\ -0.00$ \\
    78 &                     $2.98$ &                   $0.02$ &                    $44.3$ &      CC &         $4.70$ &                  $-0.61$ &                                   $-0.24$ &                                       $-0.67$ &                                 $-1.00$ \\
    79 &                     $2.96$ &                   $0.05$ &                    $44.3$ &      CC &         $4.54$ &                  $-0.62$ &                                   $-0.27$ &                                       $-0.68$ &                                 $-1.00$ \\
    80 &                     $2.98$ &                   $0.08$ &                    $44.3$ &      CC &         $4.10$ &                  $-0.66$ &                                   $-0.36$ &                                       $-0.70$ &                                 $-1.00$ \\
    81 &                     $3.05$ &                   $0.14$ &                    $44.3$ &      HR &  $1.12,\ 5.12$ &          $-0.75,\ -0.32$ &                           $-0.34,\ -0.00$ &                               $-0.87,\ -0.29$ &                         $-1.00,\ -1.00$ \\
    82 &                     $3.01$ &                   $0.20$ &                    $44.3$ &      HR &  $2.45,\ 5.89$ &          $-0.46,\ -0.22$ &                            $-0.08,\ 0.00$ &                               $-0.48,\ -0.12$ &                         $-1.00,\ -0.96$ \\
    83 &                     $3.04$ &                   $0.25$ &                    $44.3$ &      HR &  $3.20,\ 6.30$ &          $-0.29,\ -0.16$ &                             $0.00,\ 0.00$ &                               $-0.24,\ -0.06$ &                         $-1.00,\ -0.86$ \\
    84 &                     $2.95$ &                   $0.30$ &                    $44.3$ &      HR &  $3.67,\ 6.74$ &          $-0.18,\ -0.11$ &                             $0.00,\ 0.00$ &                               $-0.06,\ -0.01$ &                         $-0.98,\ -0.66$ \\
    85 &                     $2.99$ &                   $0.35$ &                    $44.3$ &      HR &  $3.89,\ 6.92$ &          $-0.14,\ -0.08$ &                             $0.00,\ 0.00$ &                               $-0.01,\ -0.00$ &                         $-0.87,\ -0.54$ \\
    86 &                     $3.05$ &                   $0.40$ &                    $44.3$ &      HR &  $3.96,\ 7.09$ &          $-0.12,\ -0.06$ &                             $0.00,\ 0.00$ &                                 $0.00,\ 0.00$ &                         $-0.80,\ -0.40$ \\
    87 &                     $2.98$ &                   $0.45$ &                    $44.3$ &      HR &  $4.07,\ 7.20$ &          $-0.10,\ -0.04$ &                             $0.00,\ 0.00$ &                                 $0.00,\ 0.00$ &                         $-0.65,\ -0.30$ \\
    88 &                     $2.98$ &                   $0.50$ &                    $44.3$ &      HR &  $4.17,\ 7.30$ &          $-0.07,\ -0.03$ &                             $0.00,\ 0.00$ &                                 $0.00,\ 0.00$ &                         $-0.50,\ -0.21$ \\
    89 &                     $3.04$ &                   $0.60$ &                    $44.3$ &      HR &  $4.35,\ 7.45$ &          $-0.03,\ -0.01$ &                             $0.00,\ 0.00$ &                                 $0.00,\ 0.00$ &                         $-0.22,\ -0.08$ \\
    90 &                     $3.00$ &                   $0.70$ &                    $44.3$ &      HR &  $4.44,\ 7.51$ &          $-0.01,\ -0.00$ &                             $0.00,\ 0.00$ &                                 $0.00,\ 0.00$ &                         $-0.09,\ -0.02$ \\
    91 &                     $2.97$ &                   $0.80$ &                    $44.3$ &      HR &  $4.48,\ 7.53$ &          $-0.00,\ -0.00$ &                             $0.00,\ 0.00$ &                                 $0.00,\ 0.00$ &                         $-0.03,\ -0.01$ \\
    92 &                     $2.99$ &                   $0.90$ &                    $44.3$ &      HR &  $4.50,\ 7.54$ &          $-0.00,\ -0.00$ &                             $0.00,\ 0.00$ &                                 $0.00,\ 0.00$ &                         $-0.01,\ -0.00$ \\
    93 &                     $2.98$ &                   $0.98$ &                    $44.3$ &      HR &  $4.50,\ 7.54$ &          $-0.00,\ -0.00$ &                             $0.00,\ 0.00$ &                                 $0.00,\ 0.00$ &                         $-0.00,\ -0.00$ \\
    94 &                     $3.96$ &                   $0.05$ &                    $44.3$ &      CC &         $0.00$ &                  $-1.00$ &                                   $-1.00$ &                                       $-1.00$ &                                 $-1.00$ \\
    95 &                     $3.96$ &                   $0.11$ &                    $44.3$ &      CC &         $2.47$ &                  $-0.80$ &                                   $-0.49$ &                                       $-0.87$ &                                 $-1.00$ \\
    96 &                     $4.01$ &                   $0.15$ &                    $44.3$ &      CC &         $3.90$ &                  $-0.68$ &                                   $-0.39$ &                                       $-0.72$ &                                 $-1.00$ \\
    97 &                     $4.01$ &                   $0.20$ &                    $44.3$ &      HR &  $1.03,\ 5.07$ &          $-0.77,\ -0.33$ &                           $-0.39,\ -0.00$ &                               $-0.88,\ -0.30$ &                         $-1.00,\ -1.00$ \\
    98 &                     $3.97$ &                   $0.25$ &                    $44.3$ &      HR &  $2.97,\ 6.04$ &          $-0.34,\ -0.20$ &                             $0.00,\ 0.00$ &                               $-0.32,\ -0.09$ &                         $-1.00,\ -0.95$ \\
    99 &                     $3.97$ &                   $0.30$ &                    $44.3$ &      HR &  $3.56,\ 6.53$ &          $-0.21,\ -0.13$ &                             $0.00,\ 0.00$ &                               $-0.10,\ -0.02$ &                         $-1.00,\ -0.79$ \\
   100 &                     $3.98$ &                   $0.35$ &                    $44.3$ &      HR &  $3.84,\ 6.78$ &          $-0.15,\ -0.10$ &                             $0.00,\ 0.00$ &                               $-0.01,\ -0.00$ &                         $-0.94,\ -0.66$ \\
   101 &                     $4.02$ &                   $0.40$ &                    $44.3$ &      HR &  $3.97,\ 6.96$ &          $-0.12,\ -0.08$ &                             $0.00,\ 0.00$ &                                 $0.00,\ 0.00$ &                         $-0.80,\ -0.51$ \\
   102 &                     $4.04$ &                   $0.45$ &                    $44.3$ &      HR &  $4.00,\ 7.07$ &          $-0.11,\ -0.06$ &                             $0.00,\ 0.00$ &                                 $0.00,\ 0.00$ &                         $-0.75,\ -0.41$ \\
   103 &                     $3.98$ &                   $0.50$ &                    $44.3$ &      HR &  $4.12,\ 7.23$ &          $-0.08,\ -0.04$ &                             $0.00,\ 0.00$ &                                 $0.00,\ 0.00$ &                         $-0.57,\ -0.27$ \\
   104 &                     $3.98$ &                   $0.60$ &                    $44.3$ &      HR &  $4.28,\ 7.40$ &          $-0.05,\ -0.02$ &                             $0.00,\ 0.00$ &                                 $0.00,\ 0.00$ &                         $-0.33,\ -0.13$ \\
   105 &                     $3.97$ &                   $0.70$ &                    $44.3$ &      HR &  $4.41,\ 7.49$ &          $-0.02,\ -0.01$ &                             $0.00,\ 0.00$ &                                 $0.00,\ 0.00$ &                         $-0.14,\ -0.04$ \\
   106 &                     $3.98$ &                   $0.80$ &                    $44.3$ &      HR &  $4.48,\ 7.52$ &          $-0.01,\ -0.00$ &                             $0.00,\ 0.00$ &                                 $0.00,\ 0.00$ &                         $-0.04,\ -0.01$ \\
   107 &                     $3.95$ &                   $0.90$ &                    $44.3$ &      HR &  $4.50,\ 7.54$ &          $-0.00,\ -0.00$ &                             $0.00,\ 0.00$ &                                 $0.00,\ 0.00$ &                         $-0.01,\ -0.00$ \\
   108 &                     $3.96$ &                   $0.98$ &                    $44.3$ &      HR &  $4.50,\ 7.54$ &          $-0.00,\ -0.00$ &                             $0.00,\ 0.00$ &                                 $0.00,\ 0.00$ &                         $-0.00,\ -0.00$ \\
   109 &                     $5.01$ &                   $0.04$ &                    $44.3$ &      CC &         $0.00$ &                  $-1.00$ &                                   $-1.00$ &                                       $-1.00$ &                                 $-1.00$ \\
   110 &                     $5.00$ &                   $0.09$ &                    $44.3$ &      CC &         $0.00$ &                  $-1.00$ &                                   $-1.00$ &                                       $-1.00$ &                                 $-1.00$ \\
   111 &                     $4.98$ &                   $0.15$ &                    $44.3$ &      CC &         $2.38$ &                  $-0.80$ &                                   $-0.49$ &                                       $-0.89$ &                                 $-1.00$ \\
   112 &                     $4.97$ &                   $0.19$ &                    $88.5$ &      CC &         $4.29$ &                  $-0.64$ &                                   $-0.38$ &                                       $-0.67$ &                                 $-1.00$ \\
   113 &                     $5.04$ &                   $0.25$ &                    $44.3$ &      HR &  $2.27,\ 5.66$ &          $-0.50,\ -0.25$ &                            $-0.01,\ 0.00$ &                               $-0.58,\ -0.17$ &                         $-1.00,\ -0.99$ \\
   114 &                     $4.96$ &                   $0.30$ &                    $44.3$ &      HR &  $3.45,\ 6.32$ &          $-0.23,\ -0.16$ &                             $0.00,\ 0.00$ &                               $-0.14,\ -0.04$ &                         $-1.00,\ -0.91$ \\
   115 &                     $5.00$ &                   $0.35$ &                    $44.3$ &      HR &  $3.83,\ 6.72$ &          $-0.15,\ -0.11$ &                             $0.00,\ 0.00$ &                               $-0.01,\ -0.00$ &                         $-0.95,\ -0.72$ \\
   116 &                     $5.00$ &                   $0.40$ &                    $44.3$ &      HR &  $3.91,\ 6.86$ &          $-0.13,\ -0.09$ &                             $0.00,\ 0.00$ &                                 $0.00,\ 0.00$ &                         $-0.88,\ -0.60$ \\
   117 &                     $4.99$ &                   $0.45$ &                    $44.3$ &      HR &  $3.98,\ 7.00$ &          $-0.12,\ -0.07$ &                             $0.00,\ 0.00$ &                                 $0.00,\ 0.00$ &                         $-0.78,\ -0.48$ \\
   118 &                     $4.98$ &                   $0.50$ &                    $44.3$ &      HR &  $4.08,\ 7.14$ &          $-0.09,\ -0.05$ &                             $0.00,\ 0.00$ &                                 $0.00,\ 0.00$ &                         $-0.63,\ -0.35$ \\
   119 &                     $5.03$ &                   $0.60$ &                    $44.3$ &      HR &  $4.23,\ 7.33$ &          $-0.06,\ -0.03$ &                             $0.00,\ 0.00$ &                                 $0.00,\ 0.00$ &                         $-0.41,\ -0.19$ \\
   120 &                     $4.95$ &                   $0.70$ &                    $44.3$ &      HR &  $4.38,\ 7.47$ &          $-0.03,\ -0.01$ &                             $0.00,\ 0.00$ &                                 $0.00,\ 0.00$ &                         $-0.18,\ -0.06$ \\
   121 &                     $5.00$ &                   $0.80$ &                    $44.3$ &      HR &  $4.47,\ 7.52$ &          $-0.01,\ -0.00$ &                             $0.00,\ 0.00$ &                                 $0.00,\ 0.00$ &                         $-0.05,\ -0.02$ \\
   122 &                     $5.00$ &                   $0.90$ &                    $44.3$ &      HR &  $4.50,\ 7.53$ &          $-0.00,\ -0.00$ &                             $0.00,\ 0.00$ &                                 $0.00,\ 0.00$ &                         $-0.01,\ -0.00$ \\
   123 &                     $5.00$ &                   $0.98$ &                    $44.3$ &      HR &  $4.50,\ 7.54$ &          $-0.00,\ -0.00$ &                             $0.00,\ 0.00$ &                                 $0.00,\ 0.00$ &                         $-0.00,\ -0.00$ \\
\enddata

\tablecomments{Here `HR', `MG', and `CC' represent \texttt{Hit \& Run}, \texttt{Merger}, and \texttt{Catastrophic} collisions, respectively. For \texttt{Merger} and \texttt{Catastrophic} collisions, we present the final mass along with the overall fractional mass change for each layer of the collision remnant. In the case of \texttt{Hit \& Run} collisions, we report the same quantities for both the lower mass and the higher mass planets, respectively, with their values separated by a comma.}

\end{deluxetable*}


\appendix
\restartappendixnumbering
\setcounter{section}{0}
\newcommand{\atabref}[1]{\hyperref[#1]{Table~\ref*{#1}}}

\section{Resolution Convergence Test}\label{app:convergence}

For the SPH simulations conducted in our study, we chose a resolution of $\sim 10^5$ SPH particles per planet. In this section, we investigate the impact of resolution in our results by varying the number of SPH particles. For this purpose, we consider three collisions with relative velocities close to the escape velocity ($\vimpact/\vesc \sim 1$) but with different impact parameters. Firstly, we examine a \texttt{Merger} scenario with $\bimpact = 0.10$. Subsequently, we study another collision with $\bimpact = 0.45$, which eventually results in a \texttt{Merger} after experiencing multiple grazing encounters. Finally, we investigate a \texttt{Hit \& Run} collision with $\bimpact = 0.60$. In each instance, we conduct two additional simulations: one with a lower resolution of $\sim 5 \times 10^4$ particles per planet and another with a higher resolution of $\sim 2 \times 10^5$ SPH particles per planet, keeping identical initial conditions. Furthermore, to assess the sensitivity of our findings to precise initial conditions, we conduct three additional simulations for each scenario, each involving slight variations in initial conditions with our adopted resolution of $\sim 10^5$ particles per planet. Detailed initial conditions and the results of these simulations are provided in \atabref{app:tab:sph_conv}.

\centerwidetable
\begin{deluxetable}{cccccccccc}
\tabletypesize{\footnotesize}
\tablecaption{Simulations conducted for resolution convergence test.}
\label{app:tab:sph_conv}

\tablehead{
    \colhead{Number} &
    \colhead{$v_{\rm{im}}/v_{\rm{esc}}$} &
    \colhead{$b\  /\ (r_{1} + r_{2})$} &
    \colhead{N$_{\rm{SPH}}$} &
    \colhead{$t_{\rm{stop}}/\rm{hour}$} &
    \colhead{Outcome} &
    \colhead{$m$} &
    \colhead{$\Delta m / m_{\rm{i}}$} &
    \colhead{$\Delta m_{\rm{mantle}} / m_{\rm{mantle,i}}$} &
    \colhead{$\Delta m_{\rm{gas}} / m_{\rm{gas,i}}$}
    }

\startdata
  1 &                    $0.979$ &      $0.100$ &  $\sim5\times 10^4$ &                 $44.3$ &      MG &        $11.04$ &                  $-0.08$ &                                        $0.00$ &                                 $-0.56$ \\
  2* &                    $0.979$ &      $0.100$ &          $\sim10^5$ &                 $44.3$ &      MG &        $11.05$ &                  $-0.08$ &                                        $0.00$ &                                 $-0.55$ \\
  3 &                    $0.979$ &      $0.100$ &  $\sim2\times 10^5$ &                 $44.3$ &      MG &        $11.00$ &                  $-0.09$ &                                        $0.00$ &                                 $-0.58$ \\
  4 &                    $1.037$ &      $0.113$ &          $\sim10^5$ &                 $44.3$ &      MG &        $11.02$ &                  $-0.08$ &                                        $0.00$ &                                 $-0.57$ \\
  5 &                    $1.023$ &      $0.114$ &          $\sim10^5$ &                 $44.3$ &      MG &        $11.00$ &                  $-0.09$ &                                        $0.00$ &                                 $-0.58$ \\
  6 &                    $0.998$ &      $0.119$ &          $\sim10^5$ &                 $44.3$ &      MG &        $11.09$ &                  $-0.08$ &                                        $0.00$ &                                 $-0.54$ \\
\hline
  7 &                    $1.000$ &      $0.449$ &  $\sim5\times 10^4$ &                $332.0$ &      MG &         $9.35$ &                  $-0.22$ &                                       $-0.12$ &                                 $-1.00$ \\
  8* &                    $1.000$ &      $0.449$ &          $\sim10^5$ &                $332.0$ &      MG &         $9.26$ &                  $-0.23$ &                                       $-0.14$ &                                 $-1.00$ \\
  9 &                    $1.000$ &      $0.449$ &  $\sim2\times 10^5$ &                $332.0$ &      MG &         $9.92$ &                  $-0.18$ &                                       $-0.04$ &                                 $-1.00$ \\
 10 &                    $0.971$ &      $0.447$ &          $\sim10^5$ &                $332.0$ &      MG &         $9.49$ &                  $-0.21$ &                                       $-0.10$ &                                 $-1.00$ \\
 11 &                    $1.007$ &      $0.447$ &          $\sim10^5$ &                $332.0$ &      MG &         $9.49$ &                  $-0.21$ &                                       $-0.10$ &                                 $-1.00$ \\
 12 &                    $1.005$ &      $0.449$ &          $\sim10^5$ &                $332.0$ &      MG &         $9.73$ &                  $-0.19$ &                                       $-0.07$ &                                 $-1.00$ \\
\hline
 13 &                    $1.041$ &      $0.600$ &  $\sim5\times 10^4$ &                 $44.3$ &      HR &  $4.36,\ 7.52$ &          $-0.03,\ -0.00$ &                                 $0.00,\ 0.00$ &                         $-0.21,\ -0.02$ \\
 14* &                    $1.041$ &      $0.600$ &          $\sim10^5$ &                 $44.3$ &      HR &  $4.35,\ 7.52$ &          $-0.03,\ -0.00$ &                                $-0.00,\ 0.00$ &                         $-0.21,\ -0.02$ \\
 15 &                    $1.041$ &      $0.600$ &  $\sim2\times 10^5$ &                 $44.3$ &      HR &  $4.35,\ 7.51$ &          $-0.03,\ -0.00$ &                                $-0.00,\ 0.00$ &                         $-0.21,\ -0.03$ \\
 16 &                    $0.977$ &      $0.596$ &          $\sim10^5$ &                 $44.3$ &      HR &  $4.32,\ 7.55$ &           $-0.04,\ 0.00$ &                                $-0.01,\ 0.01$ &                         $-0.23,\ -0.02$ \\
 17 &                    $1.039$ &      $0.599$ &          $\sim10^5$ &                 $44.3$ &      HR &  $4.38,\ 7.53$ &          $-0.03,\ -0.00$ &                                $-0.00,\ 0.00$ &                         $-0.18,\ -0.02$ \\
 18 &                    $0.958$ &      $0.600$ &          $\sim10^5$ &                 $44.3$ &      HR &  $4.33,\ 7.55$ &           $-0.04,\ 0.00$ &                                $-0.01,\ 0.01$ &                         $-0.22,\ -0.01$ \\
\enddata

\tablecomments{Here `HR' and `MG' represent \texttt{Hit \& Run} and \texttt{Merger} collisions, respectively. For \texttt{Merger} collisions, we present the final mass along with the overall fractional mass change for the mantle and the gaseous envelope of the collision remnant. In the case of \texttt{Hit \& Run} collisions, we report the same quantities for both the lower mass and the higher mass planets, respectively, with their values separated by a comma. In each of the simulations described here, the relative change in the core mass is insignificant. The simulations marked by and asterisk are used in the primary analysis of this study.}

\end{deluxetable}

\begin{figure}[htb]
\hspace*{2.2cm}\includegraphics[width=0.7\textwidth]{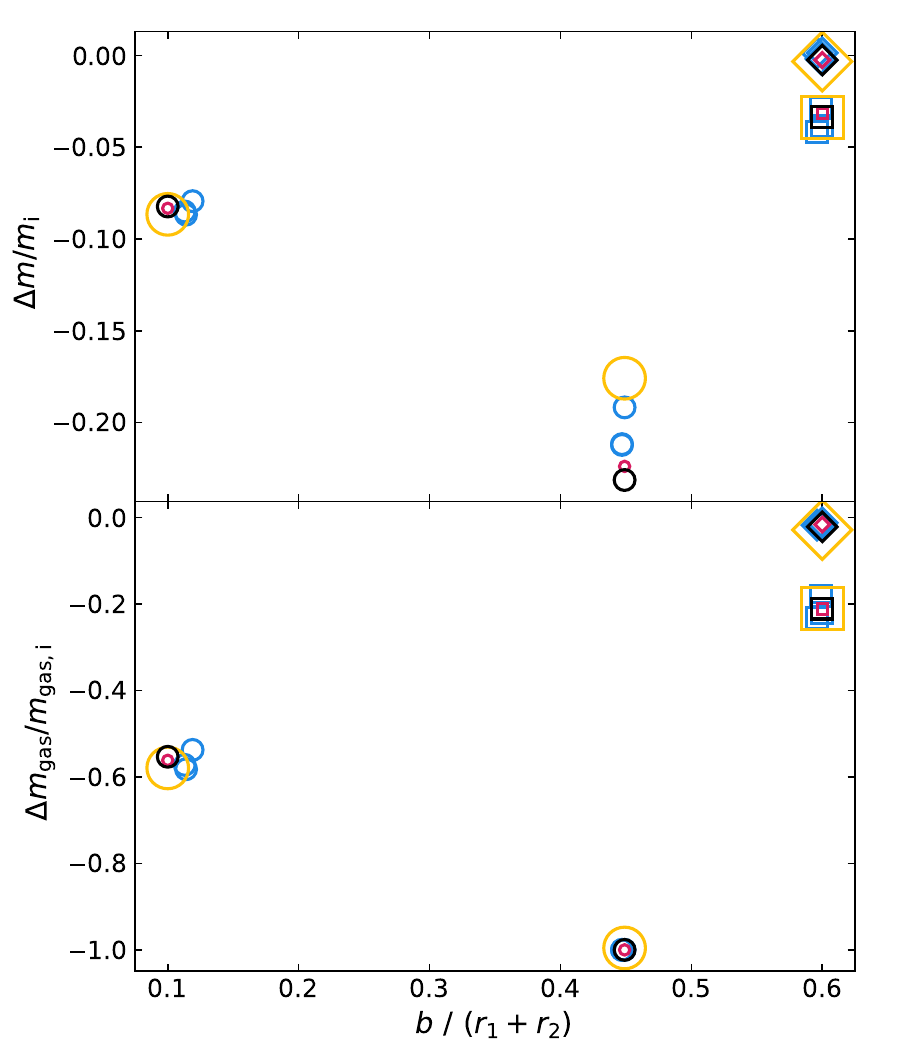}
\caption{The fractional change in the overall planet mass (top), and the atmospheric mass (bottom) as a function of the impact parameter. The circles represent collisions leading to \texttt{Merger}. Square and diamond markers represent the higher-mass and lower-mass planets for \texttt{Hit \& Run} collisions, respectively. The additional simulations conducted with a resolution of $\sim10^5$ SPH particles per planet using minor variations in the initial conditions are shown in blue markers. Red (yellow) markers represent simulations conducted with resolutions of $\sim5\times 10^4$ ($\sim2\times 10^5$) SPH particles per planet. The corresponding simulations used in our primary analysis of the study are shown in black. The marker sizes are scaled to the simulation resolutions. Some of the blue markers overlap due to identical simulation results (see \atabref{app:tab:sph_conv}). The proximity between the black and the colored markers indicates that our results are expected to be robust against minor variations of initial conditions as well as reasonable variations in the resolution of the simulations.}
\label{app-fig:convergence}
\end{figure}

In all the supplementary simulations conducted in each scenario, we find that the overall outcome of the results (i.e., \texttt{Merger} or \texttt{Hit \& Run}) have remained unchanged (\atabref{app:tab:sph_conv}). Nevertheless, there can be slight variations in the final planet mass or the atmospheric mass of the collision remnants. Overall, the results from the representative simulations with different resolutions generally remain within the expected dispersion due to tiny variations in initial conditions in those modeled with our fiducial resolution. \autoref{app-fig:convergence} shows the fractional change in planet masses and the fractional change in the H/He envelope of the planets for the simulations considered here. In the first case (\texttt{Merger} with $\bimpact \sim 0.1$), we find that the final planet mass remains consistent (within $\lesssim 1\%$) regardless of the resolution used or minor variations in initial conditions at the same resolution. The fractional loss in atmospheric mass shows $\sim4\%$ spread in the multiple simulations with the fiducial resolution. The results obtained from simulations with different resolutions are within this statistical dispersion caused by minor variations in the initial conditions. The same trend holds for the \texttt{Hit \& Run} simulations with $\bimpact \sim 0.6$. However, in the simulations with $\bimpact \sim 0.45$, we find relatively more significant differences (up to $\sim 7\%$) in the final planet masses across simulations with different resolutions as well as simulations with slightly varied initial conditions at the fiducial resolution. This relatively higher level of dispersion in final results can be attributed to the inherently violent nature of these collisions. Since these collisions are near the boundary of the \texttt{Merger} and \texttt{Hit \& Run} collisions and involve multiple grazing encounters before the eventual \texttt{Merger}, even slight variations in collision kinematics at any stage can result in relatively more significant differences in the collision remnants.

%


\bibliography{ref}{}
\bibliographystyle{aasjournal}
\end{document}